\documentclass[acmtog,anonymous=false,review=false]{acmart}
\acmSubmissionID{496}

\usepackage{booktabs} 

\usepackage{geometrycollective}
\usepackage{WalkOnStars}
\usepackage{mathdefs}

\newcounter{algo}
\newenvironment{algo}[1]
{ \refstepcounter{algo}\noindent\rule{\columnwidth}{1.25pt}\vspace{-.2\baselineskip} \\ \textbf{Algorithm~\thealgo} #1\vspace{-.55\baselineskip} \\ \noindent\rule{\columnwidth}{.5pt}\vspace{-1.2\baselineskip} }
{ \vspace{-.8\baselineskip}\noindent\rule{\columnwidth}{.5pt}\vspace{-\baselineskip} }
\algnewcommand{\LeftComment}[1]{\textcolor{commentblue}{\(\triangleright\)\textit{#1}}}

\setcopyright{rightsretained}
\acmJournal{TOG}
\acmYear{2023}
\acmVolume{42}
\acmNumber{4}
\acmArticle{1}
\acmMonth{8}
\acmPrice{15.00}
\acmDOI{10.1145/3592398}

\begin{teaserfigure}
  \centering
  \includegraphics{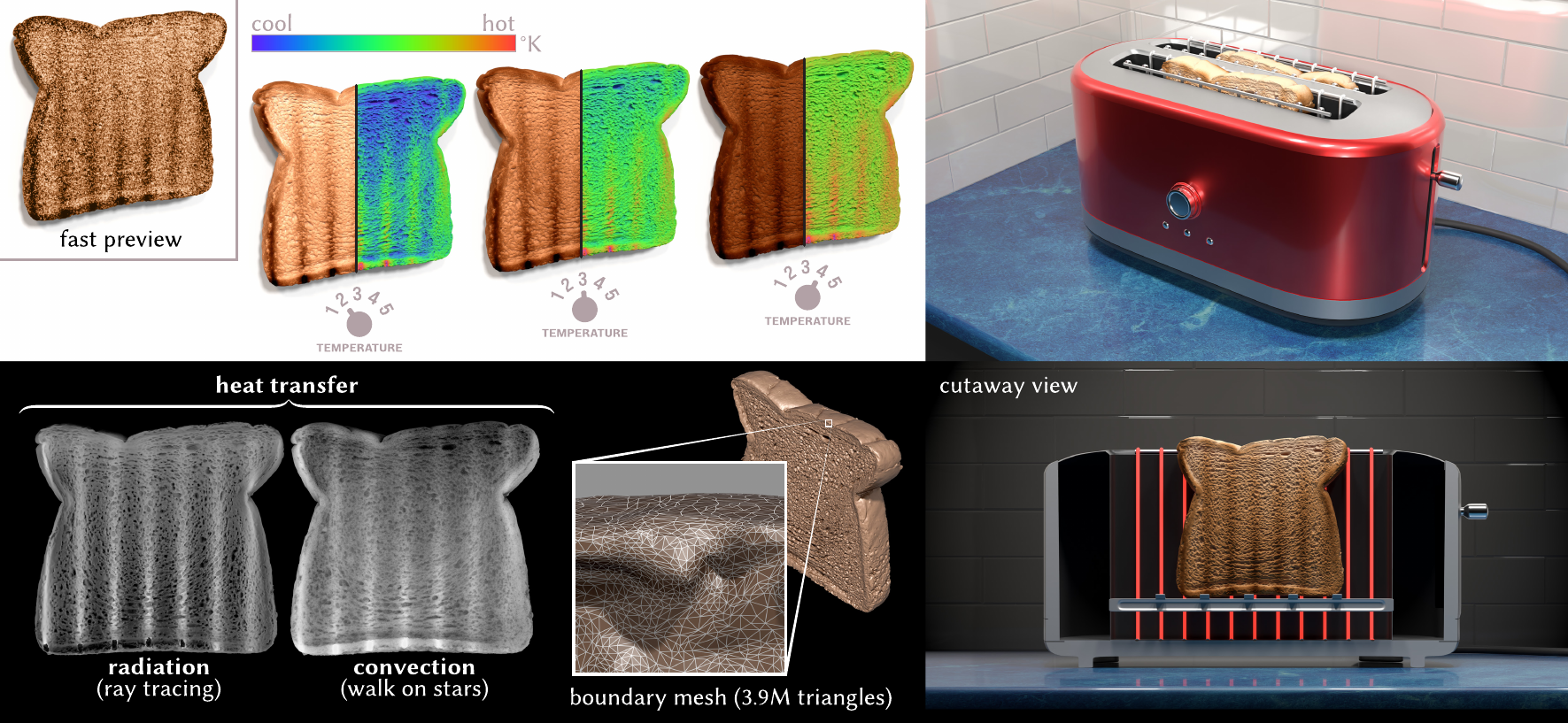}
   \caption{The walk on stars (WoSt) method handles mixed Dirichlet and Neumann boundary conditions, enabling it to model a richer class of problems than the original walk on spheres (WoS) method.  Here for instance we simulate diffusive convective heat transfer from a toaster (Dirichlet) to a piece of bread (Neumann) by solving a Laplace equation with mixed boundary conditions \figloc{(top and bottom right)}, complementing the radiative transfer computed via ray tracing \figloc{(bottom left)}.  As with ray tracing, we can simulate directly on the full high-resolution data \figloc{(bottom center)} without generating a volume mesh or forming a global stiffness matrix.  Since results are progressive, we can get a preview of how the toast will look faster than it takes to toast a real piece of bread \figloc{(top left)}.}
   \label{fig:teaser}
\end{teaserfigure}

\begin{document}
\title{Walk on Stars: A Grid-Free Monte Carlo Method for PDEs with Neumann Boundary Conditions}

\author{Rohan Sawhney}
\authornote{ and $^\dagger$ indicate equal contribution.}
\email{rohansawhney@cs.cmu.edu}
\affiliation{%
  \institution{Carnegie Mellon University and NVIDIA}
  \country{USA}
}
\orcid{0000-0002-3661-1554}

\author{Bailey Miller}
\authornotemark[1]
\email{bmmiller@andrew.cmu.edu}
\affiliation{%
  \institution{Carnegie Mellon University}
  \country{USA}
}
\orcid{0009-0009-0881-0351}

\author{Ioannis Gkioulekas}
\authornotemark[2]
\email{igkioule@cs.cmu.edu}
\affiliation{%
  \institution{Carnegie Mellon University}
  \country{USA}
}
\orcid{0000-0001-6932-4642}

\author{Keenan Crane}
\authornotemark[2]
\email{kmcrane@cs.cmu.edu}
\affiliation{%
  \institution{Carnegie Mellon University}
  \streetaddress{5000 Forbes Ave}
  \city{Pittsburgh}
  \state{PA}
  \postcode{15213}
  \country{USA}
}
\orcid{0000-0003-2772-7034}

\begin{abstract}
   Grid-free Monte Carlo methods based on the \emph{walk on spheres (WoS)} algorithm solve fundamental partial differential equations (PDEs) like the Poisson equation without discretizing the problem domain or approximating functions in a finite basis.  Such methods hence avoid aliasing in the solution, and evade the many challenges of mesh generation.  Yet for problems with complex geometry, practical grid-free methods have been largely limited to basic Dirichlet boundary conditions. We introduce the \emph{walk on stars (WoSt)} algorithm, which solves linear elliptic PDEs with arbitrary mixed Neumann and Dirichlet boundary conditions.  The key insight is that one can efficiently simulate reflecting Brownian motion (which models Neumann conditions) by replacing the balls used by WoS with \emph{star-shaped} domains.  We identify such domains via the closest point on the visibility silhouette, by simply augmenting a standard bounding volume hierarchy with normal information. Overall, WoSt is an easy modification of WoS, and retains the many attractive features of grid-free Monte Carlo methods such as progressive and view-dependent evaluation, trivial parallelization, and sublinear scaling to increasing geometric detail.
\end{abstract}

%
%


%
%


%
%
%
%

\maketitle

\section{Introduction}\label{sec:Introduction}

Systems throughout nature---and in our everyday lives---exhibit vast geometric and material complexity (Figures \ref{fig:teaser}, \ref{fig:Lungs}).  Although Monte Carlo methods have been enormously successful for photorealistic rendering, there remains a large divide between our ability to \emph{visualize} and \emph{simulate} natural phenomena.  To make complex simulation problems feasible, a common approach is to simplify the original model, \eg{}, via coarsening, homogenization, or learning.  Yet in many physical systems, subtle differences in fine-scale geometry have a major impact on large-scale behavior.  Hence, even if the end goal is to use a lower-dimensional model, one must have tools that can accurately fit such a model, starting from the original geometry.  More broadly, models of real physical systems must often integrate disparate phenomena (say, light transport and heat transfer), which classically demand very different computational tools (\figref{Lizard}).  For all these reasons, we tend to shy away from simulating the natural world at its original level of complexity, either by making gross approximations---or by tempering our ambition.

Prompted by the disparity between rendering and simulation, \citet{Sawhney:2020:MCGP} advocate the use of \emph{grid-free Monte Carlo methods} to solve partial differential equations (PDEs) on domains of extreme geometric complexity. Such methods need not discretize the problem domain (as in finite difference methods), or even pick a finite basis of functions (as in finite element and boundary element methods). Instead, like ray tracing methods, they require only pointwise access to geometry via closest point queries. This setup makes grid-free methods especially attractive in simulation scenarios where solution values or derivatives (\eg, forces) need only be evaluated at a few key points of interest, rather than densely over the entire domain. Yet grid-free Monte Carlo methods have one major shortcoming: since their development in the 1950s, they have not been extended to many PDEs beyond the original Dirichlet Laplace problem studied by \citet{Muller:1956:WOS}. In this paper we take a basic but important step forward by developing a practical strategy for incorporating \emph{Neumann boundary conditions}---which are a basic component of virtually every real physical model.

\paragraph{Basic Approach.} The original WoS method solves Dirichlet problems by simulating random walks that ultimately get absorbed into the boundary (\figref{WoSvsWoSt}, \figloc{top left}). Rather than simulate many small steps of an isotropic \emph{Brownian motion} (\figref{BMvsRBM}), this process is greatly accelerated by sampling the next point from the largest empty ball around the current point (\secref{WalkOnSpheres}). To model Neumann conditions, one must also simulate \emph{reflecting} random walks that ``bounce'' off the boundary (\figref{BMvsRBM}, \figloc{top right \& bottom})
\setlength{\columnsep}{1em}
\setlength{\intextsep}{0em}
\begin{wrapfigure}[10]{r}{57pt}
   \includegraphics[width=\linewidth]{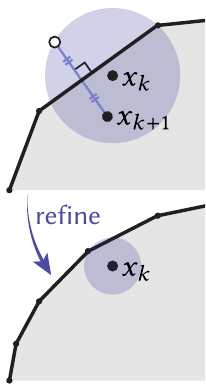}
\end{wrapfigure}
\citep{Grebenkov:2006:PRBM, Grebenkov:2007:RBM}.
In a half space, reflecting walks amount to just taking the absolute value of Brownian motion in one coordinate direction.  Hence, for polyhedral domains, a na\"{i}ve strategy for simulating reflections would be to sample the largest ball that intersects only a single boundary face, and perform a reflection across the boundary plane if the sampled point falls outside the domain.  However, the efficiency of this strategy quickly drops as the boundary mesh is refined.

Our strategy, which we call \emph{walk on stars (WoSt)}, is both more efficient and more general.  In short, we identify a large star-shaped region around the current point, and sample a point on its boundary by picking a random direction (\figref{WoSvsWoSt}, \secref{Method}). This strategy can be viewed as a Monte Carlo estimator for the \emph{boundary integral equation (BIE)} of a Laplace problem (\secref{Background}).  WoSt hence takes steps that are independent of the level of tessellation, and are typically much larger than the empty balls used by WoS (Sections \ref{sec:WoSEstimator} \& \ref{sec:MonteCarloComparisons}).  Moreover, this strategy applies to domains that are not polyhedral, and unlike past WoS-based strategies is not limited to convex domains (\secref{RandomWalkStarShapedDomains}).  The only question that must be answered is: how do we find star-shaped regions?  In this paper we propose one strategy, using the \emph{visibility silhouette}, which is easy to implement efficiently without much overhead (\secref{GeometricQueries}).  Fundamentally, however, the WoSt approach relies only on the use of star-shaped regions---not on any particular method used to compute them. Importantly, WoSt requires only few modifications to an existing WoS implementation, and achieves sublinear scaling to geometric detail using essentially the same data structures as WoS. It thus provides the same advantages as WoS (progressive evaluation, trivial parallelization, robustness to defective geometry, \etc{}), while being applicable to a broader class of problems.

\begin{figure}[t]
    \centering
    \includegraphics[width=\columnwidth]{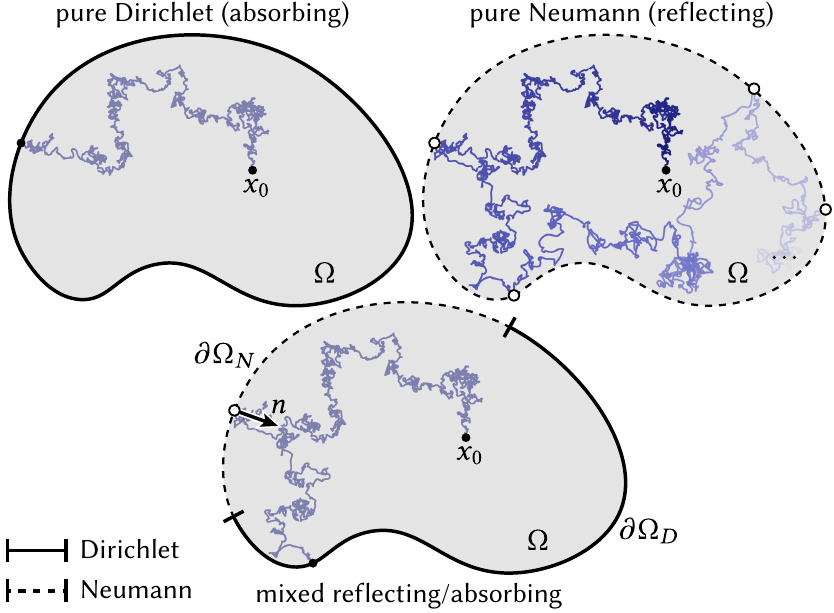}
    \caption{A Brownian random walk terminates when it hits an absorbing Dirichlet boundary $\partial \domain_D$ (\figloc{top left}), but is pushed back into the domain along the inward normal to a reflecting Neumann boundary $\partial \domain_N$ (\emph{bottom}). The walk continues forever if the boundary is only reflecting (\emph{top right}).}
    \label{fig:BMvsRBM}
\end{figure}

\begin{figure}[t]
    \centering
    \includegraphics[width=\columnwidth]{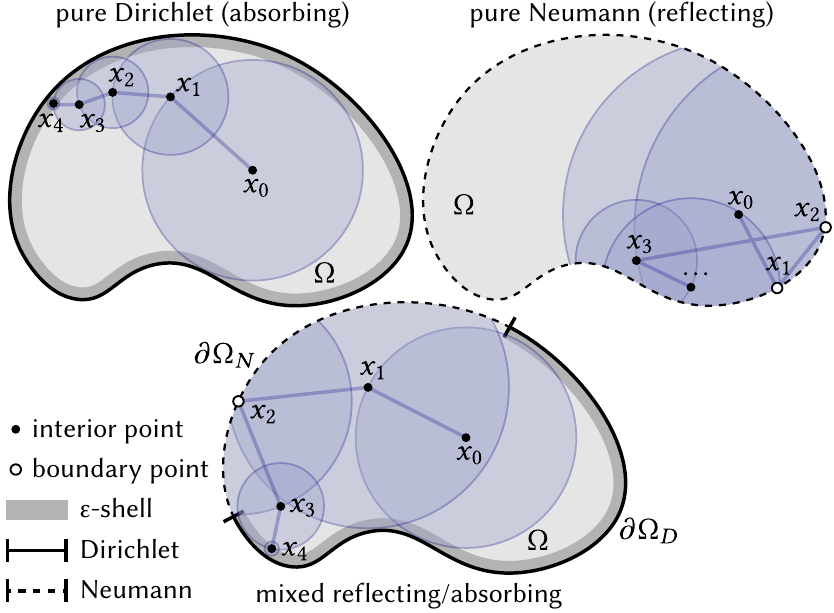}
    \caption{\emph{Top left:} Walk on spheres simulates a Brownian random walk inside an absorbing Dirichlet boundary $\partial \domain_D$, repeatedly jumping to a random point on the largest sphere centered at the current walk location. The walk is terminated when it enters an \(\varepsilon\)-shell $\partial \domain_D^{\varepsilon}$ around the boundary. \emph{Top right and bottom:} Our walk on stars algorithm generalizes WoS to domains with a reflecting Neumann boundary $\partial \domain_N$, using a sphere that can contain a subset of the reflecting boundary inside it. The next location of the walk is determined by intersecting a ray with a randomly-sampled direction against the sphere and the visible portions of $\partial \domain_N$ it contains, picking the first hit point. The walk continues forever if the boundary is only reflecting (\emph{top right}), and terminates inside $\partial \domain_D^{\varepsilon}$ otherwise (\emph{bottom}).}
    \label{fig:WoSvsWoSt}
\end{figure}

\paragraph{Limitations} For problems where boundary conditions are mostly Neumann, WoSt can take very long walks: it must reflect at the Neumann boundary, and can terminate only on the Dirichlet boundary (\figref{BMvsRBM}, \figloc{bottom}). This situation is analogous to path tracing a scene where all materials have albedo one---such as a room of perfect mirrors  (\figref{RoomOfMirrors}). Likewise, pure Neumann conditions are uncommon in many real physical scenarios---corresponding to, \eg{}, perfect insulators.  Support for more general \emph{Robin boundary conditions} would hence both improve modeling realism and increase efficiency, as more walks could terminate early (\secref{LimitationsAndFutureWork}).  In concurrent work, we also present a \emph{boundary value caching (BVC)} strategy that greatly amortizes the cost of long walks, even for Neumann-dominated problems \citep{Miller:2023:BVC}. Here we focus purely on enriching boundary conditions in grid-free Monte Carlo methods.  WoSt is otherwise limited to the same class of PDEs as WoS---namely linear elliptic PDEs such as the Poisson equation (\eqref{PoissonEquation}).  However, boundary integral formulations are readily available for the Helmholtz equation \citep[Chapter 3]{Hunter:2001:BEM}, linear elasticity \citep[Chapter 4]{Hunter:2001:BEM} and even fluids (via \emph{stochastic} integral equations) \citep{Busnello:2005:Probabilistic, Rioux-Lavoie:2022:MCFluid}).  We hence expect that WoSt will provide a valuable starting point for handling more general boundary conditions in these broader problems.

\section{Related Work}
\label{sec:RelatedWork}

We briefly discuss alternative strategies for solving PDEs; \citet[Sections 1 and 7]{Sawhney:2020:MCGP} and \citet[Section 7]{Sawhney:2022:VCWoS} describe in depth the tradeoffs between grid-free Monte Carlo and conventional, discretization-based methods like FEM, BEM, and meshless FEM, and provide extensive numerical comparisons.

\paragraph{Walk on Spheres}
Monte Carlo estimators for linear elliptic PDEs with Dirichlet boundaries, such as WoS, date back to \citet{Muller:1956:WOS}, and have recently received renewed interest following their introduction to graphics by \citet{Sawhney:2020:MCGP}. There have been rapid advances along two main thrusts: first, increasing efficiency through optimized implementation \citep{Mossberg:2021:GAM}, bidirectional formulations \citep{Qi:2022:BidirWOS}, and sample caching techniques \cite{Miller:2023:BVC}. Second, increasing generality through new estimators that can solve PDEs in infinite domains \citep{Nabizadeh:2021:Kelvin} or with variable coefficients \citep{Sawhney:2022:VCWoS}, that simulate fluid equations \citep{Rioux-Lavoie:2022:MCFluid}, and that enable differentiability for inverse problems \cite{Yilmazer:2022:DiffWOS}. Our focus is to further push the envelope along the second thrust by developing the first Monte Carlo estimator that can solve Neumann and mixed-boundary problems on general, nonconvex domains while providing a performance-bias tradeoff comparable to classic WoS.

\paragraph{Grid-based PDE Solvers}
Like WoS, WoSt does not require a volume mesh or background grid---offering critical advantages relative to grid-based methods such as finite differences and finite element methods. Namely, grid-free estimators provide output sensitivity (\ie, the ability to focus computation only on regions of interest), progressive evaluation (\ie, the ability to preview solutions as they improve), support for general geometric representations (\eg, meshes, implicit surfaces, or instanced geometry), trivial parallelization, and excellent scaling with increasing geometric detail. Whereas grid-based methods are often faster on simple models that can be easily meshed, WoSt easily handles problems whose real-world complexity places them out of reach for traditional techniques---at least not without critical sacrifices in accuracy (\figref{Lungs}). On the flip side, grid-based methods can share global information about solution values (via a coupled linear solve), whereas classic grid-free Monte Carlo methods must make independent estimates at each point---leading to fairly redundant computation.  As we show in concurrent work \cite{Miller:2023:BVC}, information sharing via sample reuse can dramatically accelerate grid-free methods as well.

\begin{figure}[t]
    \centering
    \includegraphics[width=\columnwidth]{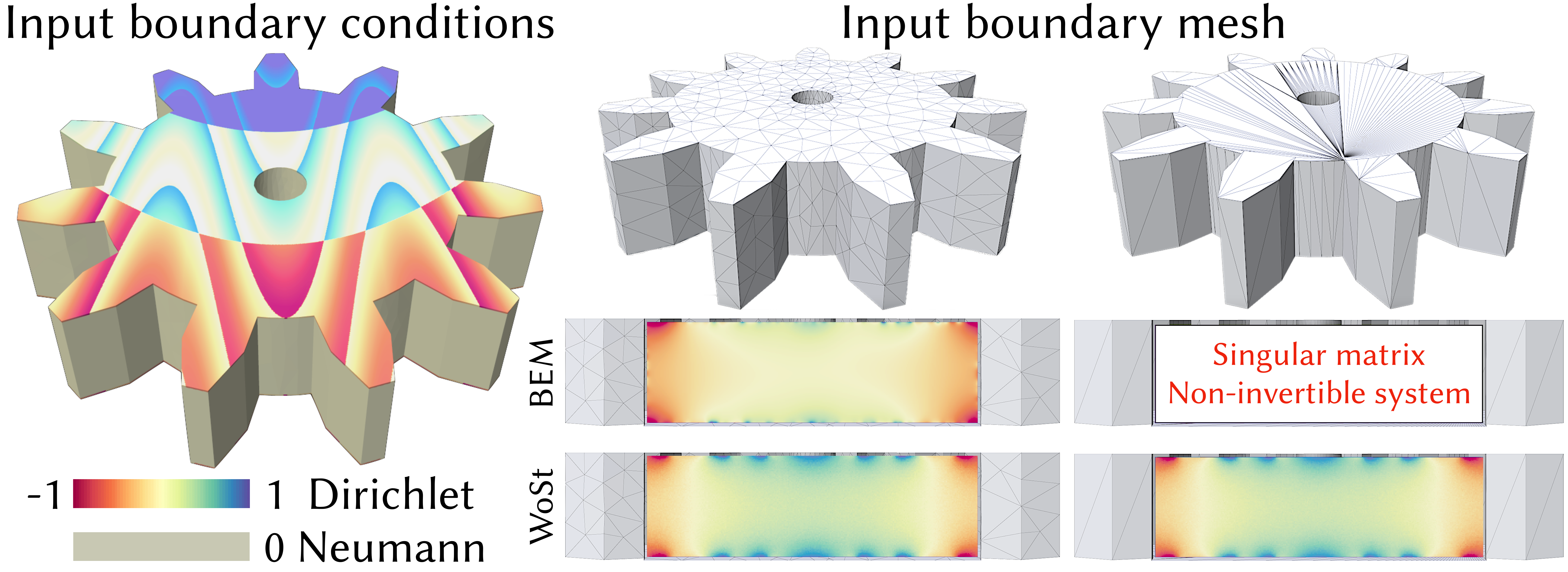}
    \caption{Finite element technology like BEM suffers from large global errors in the PDE solution without significant mesh refinement due to local aliasing of boundary data, and can fail completely on domains with irregular elements (\emph{middle row}). In contrast, our method solves PDEs without any aliasing artifacts irrespective of tesselation quality as it decouples problem inputs from the boundary representation (\emph{bottom row}). It can also handle source terms without requiring a volumetric mesh.}
    \label{fig:BEM}
\end{figure}

\begin{figure*}[t]
   \includegraphics{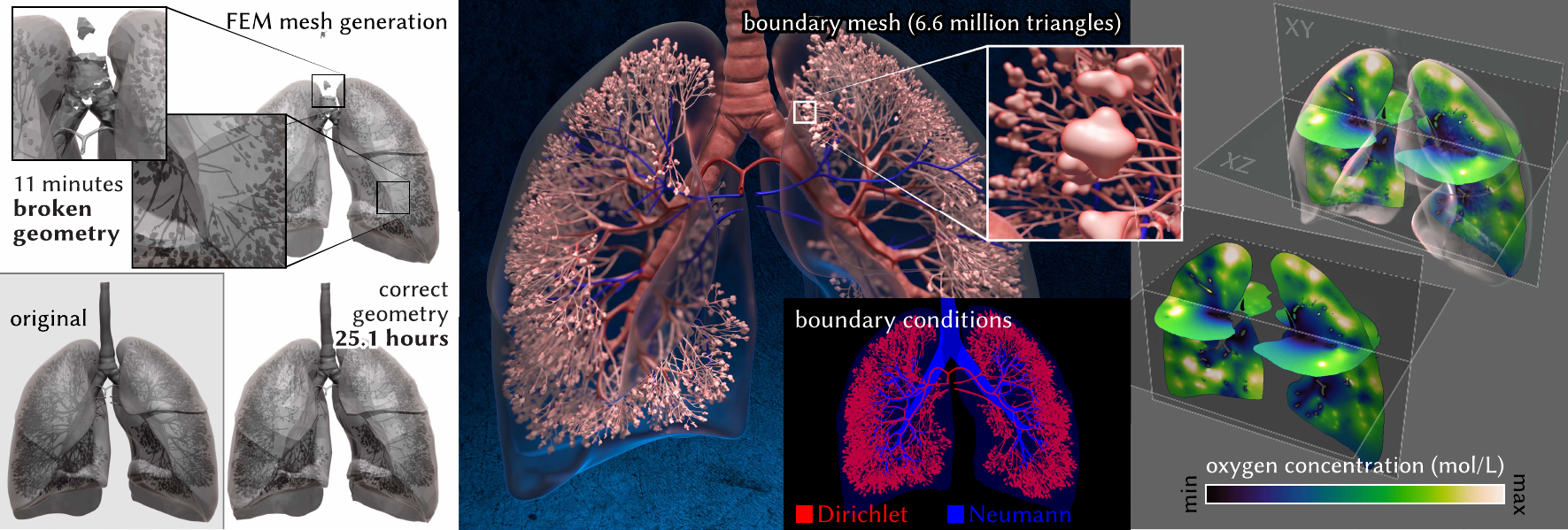}
   \caption{Where will oxygen flow at the beginning of a breath?  Here we use walk on stars to simulate gas exchange via \emph{Laplacian transport} \citep{Grebenkov:2006:PRBM}, directly on a detailed lung model with thin features \emph{(center)}.  The output-sensitivity of our method enables us to focus computation purely on the slice planes used for visualization \emph{(right)}, rather than needing to solve over the whole domain.  Attempting simulation on the same model using FEM leads to significant problems, either because meshing destroys critical details \emph{(top left)}, or takes more than 25 hours to produce a mesh that captures the original geometry \emph{(bottom left)}. In contrast, walk on stars provides near-immediate feedback that reliably reflects the true geometry and solution.}
   \label{fig:Lungs}
\end{figure*}

\paragraph{Boundary Element Methods}
Among grid-based techniques, the boundary element method (BEM) is most closely related to WoSt, as it too solves boundary integral equations (\secref{BoundaryIntegralEquation}). Whereas grid-free Monte Carlo can solve problems with volumetric data (\eg, source terms or variable PDE coefficients), BEM can do so only when coupled with a secondary solver such as FEM or radial basis function techniques \citep{Coleman:1991:EBE, Hunter:2001:BEM, Partridge:2012:DRBEM}. As a result, BEM inherits the shortcomings of the grid-based approaches discussed above. Even in the absence of volumetric terms, BEM discretizes the domain boundary to construct, then invert, a \emph{dense} linear system. Hence, computational complexity scales quadratically with geometric detail, requiring specialized techniques like \emph{hierarchical matrix approximation} to scale to large geometries~\citep{Hackbusch:2015:Hierarchical}. Finally, BEM must encode boundary conditions via the discretized boundary representation. The resulting aliasing can yield global artifacts or singular systems---and in some cases an inability to even produce a solution (\figref{BEM}).

\paragraph{Monte Carlo Methods for Neumann Problems}
Neumann conditions introduce additional complexity into random walk methods due to the need to reflect, rather than absorb, random walks at the boundary. Stochastic differential equation (SDE) integrators such as Euler-Maruyama \citep{Higham:2001:SDE} use discretized random walks, modeling reflections by projecting walks back into the domain \citep{Constantini:1998:RBM}. Such reflections introduce discretization error and greatly increase walk length (\figref{SDEWalk}), further exacerbating the challenges already associated with SDE integrators for Dirichlet problems: a fixed step size means either very slow runtimes (due to a large number of steps inside the domain) or large estimation bias (due to discretization error). Unlike SDE integrators, WoS (for pure Dirichlet problems) and WoSt (for mixed-boundary problems) use adaptive---and generally large---step sizes, avoid discrete approximation, and thus offer a much more favorable runtime-to-bias tradeoff (see \secref{MonteCarloComparisons} and \citet[Section 7.1.3]{Sawhney:2022:VCWoS}).


For pure Dirichlet problems, WoSt reduces to the standard WoS estimator (\secref{WalkOnSpheres}), \ie, it uses the largest sphere that does not intersect the boundary. Thus, walks converge quickly to the Dirichlet boundary, where they are absorbed.
For mixed-boundary problems, augmented WoS methods use finite-difference approximations of Neumann conditions \citep{Mascagni:2004:NeumannWos, Maire:2013:NeumannWos, Zhou:2017:NeumannWos}. However, near the Neumann boundary sphere sizes become very small, and walks behave much as in SDE integrators (\figref{WoSReflections}): numerous consecutive reflections result in slow runtimes (due to long walk lengths) and large estimation bias (due to accumulation of discretization error). In contrast, WoSt uses non-spherical, star-shaped regions that can contain large pieces of the Neumann boundary. Unlike WoS, it hence continues to take large steps near the Neumann boundary, while avoiding discretization. These differences translate into order-of-magnitude improvements in both runtime performance and estimation accuracy (\secref{MonteCarloComparisons}).

WoSt builds on the techniques of \citet{Simonov:2008:NeumannWos} and \citet{Ermakov:2009:WOH}, which also sample successive random walk locations on regions that can contain the Neumann boundary. However, these techniques are restricted to problems with convex Neumann boundaries. By sampling walk locations using star-shaped regions (\secref{RandomWalkStarShapedDomains}), WoSt can solve problems on general nonconvex domains, while also being significantly faster for convex ones.

\citet{Ding:2022:BIEWOS} recently proposed a hybrid WoS-BEM solver to handle Neumann problems. The hybrid nature of their solver means that it sacrifices key properties of Monte Carlo-only techniques, such as progressive evaluation, robustness to poorly tessellated surface geometry, trivial parallelization, and output sensitivity.

Finally, the walk on boundary (WoB) method is an alternative Monte Carlo approach for mixed-boundary problems. WoB recursively evaluates single and double layer potentials by tracing rays that reflect off the boundary \citep{Sabelfeld:2013:RWB}. Similar to \citet{Simonov:2008:NeumannWos} and \citet{Ermakov:2009:WOH}, this method currently works reliably only on problems with convex Neumann boundaries, suffering from high variance and bias in nonconvex domains---\secref{BranchingEstimator} explains why all three of these methods struggle with nonconvex domains; see also \secref{Convergence} for numerical experiments. In concurrent work, \citet{Sugimoto:2023:WoB} introduce WoB to the graphics community---their accessible overview and improvements to the technique likely open avenues to further, more efficient estimators that span the space between WoSt and WoB.

\paragraph{Accelerated Distance and Silhouette Queries} Much like path tracing, both WoS and WoSt require iterative applications of a basic set of geometric queries and sampling operations. WoS uses closest point queries to the Dirichlet boundary to determine random walk step sizes. These queries can be accelerated with a bounding volume hierarchy (BVH) \citep{TheEmbreedevelopers:2013:Embree, Sawhney:2021:FCPW, Krayer:2021:HPD}, achieving sublinear scaling relative to geometric detail. WoSt additionally uses \emph{closest silhouette point queries} to the Neumann boundary to define star-shaped regions. As we show in \secref{GeometricQueries}, these queries can likewise be accelerated using a BVH augmented to include orientation information for the geometry inside each node. We use the \emph{spatialized normal cone hierarchy (SNCH)} of \citet{Johnson:2001:SNCH}, which has previously been used to accelerate culling of back-facing geometry, local minimum distance queries, and \emph{next-event estimation} (NEE) for rendering scenes with many lights \citep{Estevez:2018:Importance}. More recently, silhouette queries have become a critical subroutine for differentiable Monte Carlo rendering \citep{Li:2018:Edge}, where they are used to estimate radiometric integrals arising from visibility discontinuities. Thus, acceleration schemes for silhouette queries in these emerging rendering algorithms are likely to benefit WoSt, and vice versa.

\section{Background}\label{sec:Background}


We first review linear elliptic PDEs and their boundary integral equation (BIE) representation, which serves as the starting point for our WoSt estimator in \secref{Method}. We also demonstrate how to derive the WoS estimator from the BIE, to aid comparison with WoSt.

\subsection{Notation}\label{sec:Notation}
For any set $\genset \subset \R^N$, we use $\partial \genset$ and $|\genset|$ for its boundary and volume, \resp{}\ We use $p^\genset$ to denote a probability density function on $\genset$, and write $x \sim p^\genset$ for a random point $x \in A$ drawn from $p^\genset$. For any point $z \in \partial \genset$, we use $n_z$ for the unit outward normal at $z$.
We say \(\partial A\) is \emph{convex} at \(z\) if all principal curvatures are positive (relative to \(n_z\)), and nonconvex otherwise.
For any point $x \in \genset$, $\closest{x}(\partial \genset) \coloneqq \argmin_{y \in \partial \genset} \norm{y - x}$ is the closest point to $x$ on $\partial \genset$, and $\ball(x, r)$ is a ball with center $x$ and radius $r$. We drop arguments for $\closest{x}$ and $\ball$ when the context is clear. We define the \emph{\(\varepsilon\)-shell} around $\partial \genset$ as $\partial \genset^{\varepsilon} \coloneqq \{x \in \domain : \norm{x - \closest{x}(\partial \genset)} < \varepsilon\}$.
Finally, we use $\Delta$ for the negative-semidefinite Laplace operator on $\R^N$.

\subsection{Linear Elliptic Equations}\label{sec:PoissonEquation}
Linear elliptic partial differential equations have broad utility in geometry processing, simulation, graphics, and scientific computing in general.  A prototypical example is the \emph{Poisson equation}
\begin{equation}
    \label{eq:PoissonEquation}
    \begin{array}{rcll}
        \Delta u(x) &=& f(x)    & \text{ on } \domain,\\
        u(x) &=& g(x) & \text{ on } \partial \domain_D,\\
        \frac{\partial u(x)}{\partial n_x} &=& h(x) & \text{ on } \partial \domain_N,
    \end{array}
\end{equation}
which describes, \eg{}, the steady-state temperature distribution over a domain \(\Omega \subset \R^N\).  Here $u: \domain \to \R$ is the unknown solution, and $f: \domain \to \R$ is a \emph{source term}, analogous to a heat source or sink. We partition the boundary $\partial \domain$ into a subset $\partial \domain_D$ where the solution has \emph{Dirichlet boundary conditions}, \ie, prescribed values \(g: \partial \domain_D \to \R\), and a subset \(\partial \domain_N\) where it has \emph{Neumann boundary conditions}, \ie, prescribed derivatives \(h: \partial \domain_N \to \R\) in the normal direction \(n\).  Either subset can be empty, in which case we say the equation has \emph{pure} Dirichlet or Neumann conditions. A function $u$ is \emph{harmonic} if it satisfies \eqref{PoissonEquation} for $f = 0$, \ie, if \(\Delta u = 0\) on \(\Omega\).

A \emph{screened Poisson equation} adds a constant \emph{absorption term} \(\sigma u\), \(\sigma \in \R_{> 0}\), modeling a medium that dampens or ``cools'' the solution:
\begin{equation}
    \label{eq:ScreenedPoissonEquation}
    \begin{array}{rcll}
        \Delta u(x) - \sigma u(x) &=& f(x)    & \text{ on } \domain,\\
    \end{array}
\end{equation}
subject to the same boundary conditions as in \eqref{PoissonEquation}. One can form more general linear elliptic equations by adding variable diffusion, drift and absorption coefficients to \eqref{ScreenedPoissonEquation}---see \citet[Section 2.2]{Sawhney:2022:VCWoS} for a detailed exposition.  For simplicity our exposition focuses on the (screened) Poisson equation, though much of the material presented here applies to more general linear PDEs (see \secref{LimitationsAndFutureWork} for further discussion).

\subsubsection{Green's Function}
\label{sec:Green'sFunction}

A \emph{Green's function} captures the influence of the source term \(f\)---in particular, it describes the solution when the source is a Dirac delta distribution \(\delta_x\) centered at a single point \(x \in \Omega\) \citep{Evans:1998:PDE, Duffy:2015:Green}.  For instance, in the case of \eqref{PoissonEquation} the Green's function \(G^\Omega(x,y)\) is the solution to the Poisson equation \(\Delta u(y) = \delta_x(y)\).  In general, a Green's function will depend on the shape of the domain \(\Omega\) and the choice of boundary conditions. Typically, Green's functions are not available in closed-form---however, explicit expressions are available for important special cases, \eg, the \emph{free-space Green's function} $\smash{\green^{\mathbb{R}^N}}$ on $\domain = \R^N$, and the Green's function $\green^B$ for a ball $\domain = \ball$ with zero-Dirichlet boundary conditions (see \appref{GreensFnsPoisson}).  The WoS and WoSt methods effectively provide a bridge between closed-form Green's functions on special domains, and solutions to PDEs on more general domains.

\subsubsection{Poisson Kernel}
\label{sec:PoissonKernel}

The \emph{Poisson kernel} likewise captures the influence of the boundary conditions on the solution, \eg{}, when the function \(g\) is a Dirac delta distribution \(\delta_x\) centered on a single boundary point \(x \in \partial\Omega\).  At any point $y \in \Omega$ with associated normal $n_y$, it can be expressed as the normal derivative of a Green's function:
\begin{equation}
    \label{eq:PoissonKernel}
    \poisson^\domain(x, y) \coloneqq \frac{\partial \green^{\domain}(x, y)}{\partial n_y}.
\end{equation}
As with Green's functions, common Poisson kernels are known explicitly in free space and for a ball (see \appref{GreensFnsScreenedPoisson}).

\subsection{Boundary Integral Equation}
\label{sec:BoundaryIntegralEquation}

The solution \(u\) to a linear PDE can be expressed via a \emph{boundary integral} involving the associated Green's function and Poisson kernel.  Assume for now that $\domain$ is a watertight domain with smooth boundary \(\partial\Omega\), and let $x$ be an evaluation point on the interior of $\domain$. We first multiply the Poisson equation in \eqref{PoissonEquation} with its Green's function $\green^{\domain}$ and integrate over $\domain$ to get
\begin{equation}
    \label{eq:BieDerivationFirst}
    0 = \int_{\domain} \green^\domain(x, y)\ \Delta u(y)\ud y\ -\ \int_{\domain} \green^\domain(x, y)\ f(y)\ud y.
\end{equation}
Applying integration by parts to the first integral, we have
\begin{align}
    \label{eq:BieDerivationSecond}
    0 &= \int_{\partial \domain} \green^\domain(x, z)\ \frac{\partial u(z)}{\partial n_z}\ud z\ -\ \int_{\domain} \nabla \green^\domain(x, y)\ \cdot \nabla u(y)\ud y\nonumber\\
      &-\ \int_{\domain} \green^\domain(x, y)\ f(y)\ud y.
\end{align}
Applying integration by parts again to the second integral in \eqref{BieDerivationSecond} and rearranging terms then yields
\begin{align}
    \label{eq:BieDerivationThird}
    \int_{\domain} u(y)\ \Delta \green^\domain(x, y) \ud y &= \int_{\partial \domain} \poisson^\domain(x,z)\ u(z)\ -\ \green^\domain(x, z)\ \frac{\partial u(z)}{\partial n_z}\ud z\nonumber\\
                                           &+ \int_{\domain} \green^\domain(x, y)\ f(y)\ud y.
\end{align}
From the definition $\Delta \green^{\domain}(x, y) = \delta^{\domain}_x(y)$ we arrive at
\begin{align}
    \label{eq:BoundaryIntegralEquationSimple}
    u(x) &= \int_{\partial \domain} \poisson^\domain(x,z)\ u(z)\ -\ \green^\domain(x, z)\ \frac{\partial u(z)}{\partial n_z}\ud z\nonumber\\
         &+ \int_{\domain} \green^\domain(x, y)\ f(y)\ud y.
\end{align}
This equation determines the solution $u$ at $x$ entirely through the solution values $u(z)$ and normal derivatives $\nicefrac{\partial u(z)}{\partial n_z}$ on the boundary $\partial \domain$, and the source values $f(y)$ inside the domain $\domain$. From \eqref{PoissonEquation}, the Dirichlet and Neumann parts of the boundary have prescribed values $g$ and $h$, \resp{}, while $f$ is specified inside the domain. To use \eqref{BoundaryIntegralEquationSimple}, we must then determine unknown solution values $u(z)$ on the Neumann boundary $\partial \domain_N$, and unknown derivative values $\nicefrac{\partial u(z)}{\partial n_z}$ on the Dirichlet boundary $\partial \domain_D$.

\subsubsection{General Setting} In practice, \eqref{BoundaryIntegralEquationSimple} cannot be used directly since the Green's function and Poisson kernel for an arbitrary domain $\domain$ are unknown. Fortunately, this equation can be generalized to the \emph{boundary integral equation} (BIE) \citep[Section 2]{Costabel:1987:BEM} where these functions are no longer tied to the domain \(\Omega\).  Instead one may use, \eg{}, the closed-form Green's function and Poisson kernel for a ball, or for \(\R^N\). Moreover, while we ultimately seek a solution on $\domain$, the BIE applies to arbitrary subdomains in $\domain$:

\begin{myTitledBox}{Boundary Integral Equation}
    For any two sets $\genset, \gensetalt \subset \domain$, and for any point $x \in \R^N$, the solution to the Poisson equation in \eqref{PoissonEquation} satisfies:
    \begin{align}
            \diffeqnSplit{,}{%
                \alpha(x)\ u(x)
            }{%
                \int_{\partial \genset} \poisson^{\gensetalt}(x, z)\ u(z)\ -\ \green^{\gensetalt}(x, z)\ \frac{\partial u(z)}{\partial n_z}\ud z
            }{%
                \int_{\genset} \green^{\gensetalt}(x, y)\ f(y)\ud y
            }\label{eq:BoundaryIntegralEquationSingleSided}
    \end{align}
    where
    \begin{equation}
        \label{eq:Alpha}
        \alpha(x) \coloneqq
        \begin{cases}
            1, \!\!& x \in \genset, \\
            1/2, \!\!& x \in \partial \genset, \\
            0, \!\!& x \notin \genset.
        \end{cases}
    \end{equation}
\end{myTitledBox}

\citet[Chapter 3.3]{Hunter:2001:BEM} provide a derivation. For screened Poisson equations we instead use the Green's function and Poisson kernel from \appref{GreensFnsScreenedPoisson}.  The BIE can be extended further---for instance, if \(\partial\genset\) is a non-smooth curve in the plane, then \(\alpha = 1-\theta/2\pi\) at a corner with interior angle \(\theta\).  Likewise, in \appref{WalkOnStarsDoubleSided} we generalize the BIE to double-sided boundary conditions.  To keep things simple, we will assume in \secref{Method} that \(\partial \genset\) is smooth, letting $\alpha = 1/2$ at all boundary points.

Both BEM and WoS can be interpreted as methods for solving the BIE, for different choices of sets \(A\) and \(C\).  Throughout we highlight boundary terms (in blue) and interior terms (in gray) to make the correspondence with \eqref{BoundaryIntegralEquationSingleSided} clear.

\subsubsection{Boundary Element Method} BEM integrates \eqref{BoundaryIntegralEquationSingleSided} over the PDE domain (\(\genset = \domain\)) using free-space kernels (\(\gensetalt = \R^N\)).  BEM does not directly support source terms \(f\), leading to the integral
\begin{equation}
    \label{eq:BoundaryIntegralEquationBEM}
    \alpha(x)\ u(x) =\; \mymathbox{intBlueL}{boundary}{\int_{\partial \domain} \poisson^{\R^N}(x, z)\ u(z)\ -\ \green^{\R^N}(x, z)\ \frac{\partial u(z)}{\partial n_z}\ud z} \;.
\end{equation}
To determine the unknown data $u$ on $\partial \domain_N$ and $\nicefrac{\partial u}{\partial n}$ on $\partial \domain_D$, BEM uses a finite basis of functions (associated with mesh nodes on a discretized boundary) to solve a dense linear system---resulting in the tradeoffs discussed in \secref{RelatedWork}.

\subsubsection{Walk on Spheres} WoS instead integrates the BIE over a ball \(\ball(x,r) \subset \domain\) centered at \(x\), adopting kernels from the ball (\(\genset = \gensetalt = \ball(x,r)\)).  At points \(z \in \partial \ball\), these kernels simplify to $\green^{\ball}(x, z) = 0$ and $\poisson^{\ball}(x, z) = \nicefrac{1}{|\partial \ball|}$ (\appref{GreensFnsPoisson}), yielding an integral
\begin{equation}
    \label{eq:WalkOnSpheresIntegral}
        \diffeqn{.}{%
            u(x)
        }{%
            \frac{1}{|\partial \ball(x, r)|} \int_{\partial \ball(x, r)}\!\!\!\!\!\!\!\! u(z) \ud z
        }{%
        \int_{\ball(x, r)}\!\!\!\!\!\!\!\! \green^{\ball}(x, y)\ f(y)\ud y
        }
\end{equation}
This setup greatly simplifies the BIE by eliminating dependence on $\nicefrac{\partial u}{\partial n}$ (hence avoiding the need for any branching estimates).  Unlike BEM, the source term \(f\) is accounted for, and one does not need to discretize the domain boundary \(\partial \domain\) nor solve a global system of equations.  Instead, the solution is evaluated by recursively using \eqref{WalkOnSpheresIntegral} to estimate \(u(z)\) on the ball boundary \(\partial \ball\), leading to the WoS algorithm (\secref{WoSEstimator}).

\subsection{Walk on Spheres Estimator}
\label{sec:WalkOnSpheres}

WoS is a \emph{Monte Carlo estimator} for \eqref{WalkOnSpheresIntegral}, which means that it approximates the boundary and interior integrals using random samples of the integrands. WoSt follows the same basic recipe, but for a different version of the BIE.  Before discussing these estimators in detail, we first review Monte Carlo integration \citep{fishman2006monte}, which is the numerical foundation for both methods.

\subsubsection{Monte Carlo Integration}\label{sec:MCIntegration}
For an $L^1$-integrable function $\phi: \genset \to \R$, the Monte Carlo method approximates the integral
\begin{align}
    I \coloneqq \int_{\genset} \phi(x)\ud x
\end{align}
using the sum
\begin{align}\label{eq:MonteCarloEstimator}
   \widehat{I}_N \coloneqq \frac{1}{N} \sum_{n=1}^{N} \frac{\phi(x_n)}{p^\genset(x_n)}, \quad x_n \sim p^\genset,
\end{align}
where $x_n$ are independent samples randomly drawn from a probability density $p^\genset$ that is nonzero on the support of $\phi$. An estimator is \emph{unbiased} if its expected value equals the true value, $\expectation{\widehat{I}_N} = I$. We can quantify the accuracy of an estimator using its expected squared error $\expectation{(\widehat{I}_N - I)^2}$, which for an unbiased estimator equals its \emph{variance} $\Var[\widehat{I}_N] \coloneqq \expectation{(\widehat{I}_N - \expectation{\widehat{I}_N})^2}$. Assuming independent samples $x_n$, variance goes to zero at a rate $O(\nicefrac{1}{N})$ \citep[Chapter 13]{Pharr:2016:PBR}. We express all estimators in this paper as \emph{single-sample estimates} $\widehat{I}$, dropping the subscript $N = 1$ for brevity. Averaging these estimates over many trials improves accuracy.

\subsubsection{WoS Estimator}
\label{sec:WoSEstimator}

A recursive single-sample estimate for \eqref{WalkOnSpheresIntegral} at a point \(x_k \in \domain\) is given by
\begin{equation}
    \label{eq:WalkOnSpheresEstimator}
        \diffeqnOp{,}{%
            \widehat{u}(x_k)\!\!
        }{%
            \!\!\frac{1}{|\partial \ball(x_k)|}\frac{\widehat{u}(x_{k+1})}{p^{\partial \ball(x_k)}(x_{k+1})}\!\!\!
        }{%
            \!\!\!\frac{\green^{\ball(x_k)}(x_{k}, y_{k+1})f(y_{k+1})}{p^{\ball(x_k)}(y_{k+1})}\!\!
        }{\coloneqq}
\end{equation}
where $x_{k+1} \in \partial \ball$ and ${y}_{k+1} \in \ball$ are sampled from probability densities $p^{\partial \ball}$ and $p^{\ball}$, \resp{}\ Typical choices are the uniform density $p^{\partial \ball} \coloneqq 1/|\partial \ball|$, and the normalized Green's function $p^{\ball} \coloneqq \green^{\ball}(x_k, y_{k+1}) / |\green^{\ball}(x_k)|$, where $|\green^{\ball}(x_k)|$ is the integral of $\green^{\ball}$ over \(\ball(x_k)\) (see \appref{GreensFns} and \citet[Section 4.2]{Sawhney:2020:MCGP}).  Recursive evaluation of \eqref{WalkOnSpheresEstimator} determines a \emph{random walk} on the points $x_0 \to x_1 \to \dots$, where each point \(x_k\) sits on a sphere centered at the previous point \(x_{k-1}\)---hence the name \emph{walk on spheres} (\figref{WoSvsWoSt}).  This walk terminates if $u(x_k)$ is known; otherwise, the process repeats.  To take big steps, WoS typically uses the largest sphere centered at $x_k$ and contained entirely in $\domain$ (\ie, $r = \norm{x-\closest{x}}$).

\subsubsection{Boundary Conditions for WoS}
\label{sec:BoundaryConditionsforWoS}

How a walk terminates depends on the particular boundary conditions, enumerated below.

\setlength{\columnsep}{1em}
\setlength{\intextsep}{0em}
\begin{wrapfigure}[7]{r}{70pt}
    \centering
    \includegraphics[width=\linewidth]{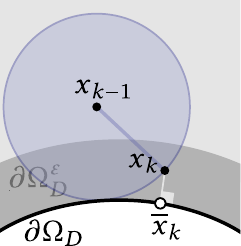}
    \label{fig:EpsilonShell}
\end{wrapfigure}%
\paragraph{Pure Dirichlet Conditions} For pure Dirichlet conditions ($\partial \domain = \partial \domain_D$), walks terminate in the \(\varepsilon\)-shell $\partial \domain^{\varepsilon}_D$ and the solution estimate is set to the boundary value \(g\) at the closest point $\closest{x}_k$, \ie, $\widehat{u}(x_k) \coloneqq g(\closest{x}_k)$. Terminating walks in the shell introduces negligible bias of order $O(\nicefrac{1}{\log \varepsilon})$ \citep{Binder:2012:WosRate}; \citet[Figure 14]{Sawhney:2020:MCGP} show experimentally that shrinking $\varepsilon$ has little impact on runtime cost. WoSt likewise uses an \(\varepsilon\)-shell to absorb walks near the Dirichlet boundary, reducing to WoS for pure Dirichlet problems.

\begin{figure}[t]
    \centering
    \includegraphics[width=\columnwidth]{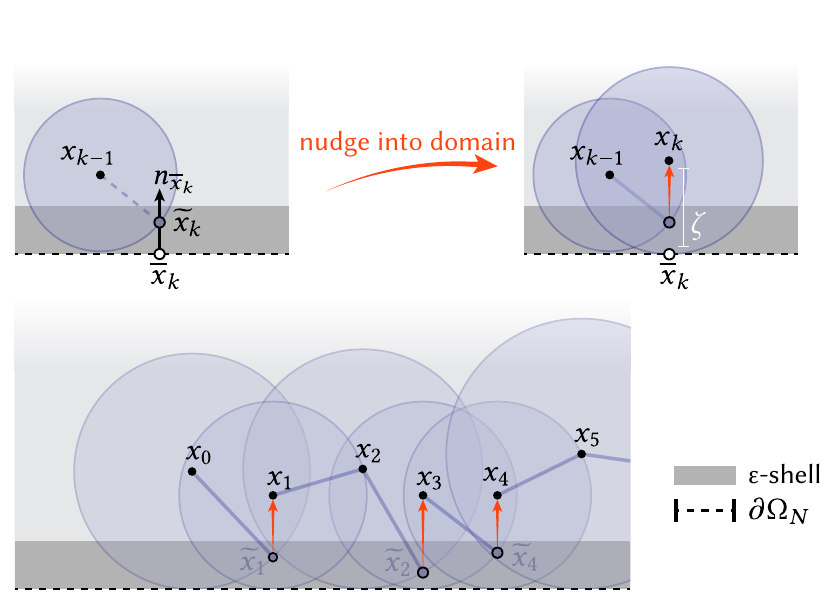}
    \caption{To simulate reflecting random walks with WoS, a standard approach \citep{Mascagni:2004:NeumannWos, Maire:2013:NeumannWos} is to offset a walk that approaches the Neumann boundary back into the domain by a fixed distance along the inward normal $n$ to the boundary (\emph{top}). This approach introduces discretization error into the reflecting walk simulation. Moreover, the resulting walks have a tendency to cling to the boundary as they are naturally attracted to it, leading to long walk lengths (\emph{bottom}).}
    \label{fig:WoSReflections}
\end{figure}

\paragraph{Mixed Dirichlet-Neumann Conditions} The standard WoS approach for mixed boundary problems \citep{Mascagni:2004:NeumannWos} also performs a random walk as above, again terminating on $\partial \domain_D^{\varepsilon}$. If the walk ever reaches a point $\widetilde{x}_k$ in the \(\varepsilon\)-shell $\partial \domain_N^{\varepsilon}$ around the Neumann boundary, then the Neumann value \(h\) at the closest point $\closest{x}_k \in \partial \domain_N$ is approximated via finite differences, \eg,
\begin{equation}
    \label{eq:FiniteDifferenceApprox}
    h(\closest{x}_k) \approx \frac{u(\closest{x}_k + \zeta n_{\closest{x}_k}) - u(\closest{x}_k)}{\zeta},
\end{equation}
where \(\zeta > \varepsilon\) is a constant.  The solution estimate at $x_k$ is then
\begin{equation}
    \label{eq:Reflection}
    \widehat{u}(x_k) \coloneqq \widehat{u}(\closest{x}_k + \zeta n_{\closest{x}_k}) - \zeta h(\closest{x}_k) \approx u(\closest{x}_k).
\end{equation}
In other words, $-\zeta h(\closest{x}_k)$ is added to the running estimate, and the walk continues as usual from the point \(\closest{x}_k + \zeta n_{\closest{x}_k}\) obtained by nudging \(\closest{x}_k\) back into the domain by a fixed distance \(\zeta\) along the inward unit normal (\figref{WoSReflections}, \emph{top}). \citet{Mascagni:2004:NeumannWos} call this procedure a \emph{boundary reflection}; \citet{Maire:2013:NeumannWos} and \citet{Zhou:2017:NeumannWos} provide more sophisticated approximations using higher-order differences.

Unfortunately, such reflections are often impractical for problems with a large Neumann boundary $\partial \domain_N$: the finite difference approximation introduces significant bias if $\zeta$ is much larger than \(\varepsilon\)--yet if $\zeta$ is only slightly larger than $\varepsilon$, random walks ``stick'' to $\partial \domain_N^{\varepsilon}$, taking many small steps before escaping toward the interior (\figref{WoSReflections}, \emph{bottom}). \figref{McComparisons} shows that, in practice, boundary reflections yield both slow runtime and large accumulated bias. WoSt avoids these issues by considering larger spheres that contain the Neumann boundary (\secref{Method}), greatly improving both accuracy and efficiency.

\paragraph{Pure Neumann Conditions} The solution to a Poisson equation with pure Neumann boundary conditions is determined only up to an additive constant.  From the random walk perspective, there is no Dirichlet boundary to terminate on, hence contributions from $h$ and $f$ accumulate forever. However, shorter walks tend to resolve high-frequency details in the solution, whereas the contribution from independent longer walks is more spatially uniform (see \figref{PureNeumannConstant}). Based on this observation, \citet{Maire:2013:NeumannWos} describe a WoS estimator that stops the simulation once walks become longer than a certain length, pinning an additive constant to the solution. For WoSt we instead apply \emph{Tikhonov regularization}, which makes the solution unique by adding a small absorption term \(\sigma u\) to the PDE (resulting in a screened Poisson equation). In particular, we switch to this PDE when a walk gets longer than a user-specified length (\figref{Tikhonov}), which adds a small but controlled amount of bias. We then terminate walks via \emph{Russian roulette} \citep[Section 13.7]{Pharr:2016:PBR}, using a termination probability proportional to the Poisson kernel of a screened Poisson equation (\eqref{ScreenedPoissonKernelsSphere}).

\section{The Walk on Stars Algorithm}\label{sec:Method}

In this section, we develop \emph{walk on stars} (WoSt), a recursive estimator for mixed boundary-value problems like \eqref{PoissonEquation}. Like WoS, WoSt takes large steps inside the domain $\domain$ to quickly reach the Dirichlet boundary---yet unlike WoS, WoSt can also take large steps near the Neumann boundary without incurring large bias. Our method is still entirely grid-free, \ie, neither \(\Omega\) nor \(\partial\Omega\) has to be discretized; we need only query the boundary geometry \(\partial \domain\) via ray intersections and modified distance queries (\secref{GeometricQueries}). Here we assume for simplicity that the domain $\domain$ is a compact subset of \(\R^N\); see \appref{WalkOnStarsDoubleSided} for extensions to open domains and double-sided boundary conditions.  Detailed pseudocode is provided in \algref{WalkOnStars}.

\subsection{Star-Shaped Subdomains}
\label{sec:StarShapedSubdomains}

\newcommand{\BallContainingNeumannBoundary}{%
\setlength{\columnsep}{1em}
\setlength{\intextsep}{0em}
\begin{wrapfigure}[9]{r}{81pt}
    \centering
    \includegraphics{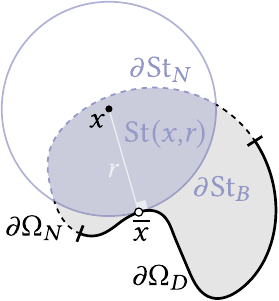}
    \label{fig:SimonovSubDomain}
\end{wrapfigure}
}

\BallContainingNeumannBoundary{} In lieu of balls, WoSt considers regions that are \emph{star-shaped} with respect to a point $x$ \citep{hansen2020starshaped}: any ray cast from $x$ must intersect the region boundary only once.  Though in principle any star-shaped subdomain could be used, we use regions \(\ssd(x, r)\) given by the component of \(\ball(x, r)\ \cap\ \domain\) containing \(x\), for a particular choice of ball \(\ball\) (see \secref{SamplingStarShapedRegions}). Similar to \eqref{PoissonEquation}, we partition the region boundary into a Neumann part $\ssdboundary_N \coloneqq \partial \Omega_N \cap \ssdboundary$ with prescribed normal derivatives $\nicefrac{\partial u}{\partial n} = h$ from \eqref{PoissonEquation}, and a spherical part $\ssdboundary_{\ball} \coloneqq \partial \ball \cap \ssdboundary$ (see inset).

\subsection{Boundary Integral Formulation}
\label{sec:BieWoSt}

Letting $\genset \coloneqq \ssd$ and $\gensetalt \coloneqq B$ in \eqref{BoundaryIntegralEquationSingleSided}, the BIE becomes
\begin{align}
        \diffeqnSplit{.}{%
            \alpha(x)u(x)
        }{%
            \int_{\ssdboundary(x, r)}\!\!\!\!\!\!\!\!\!\!\!\!\poisson^{\ball}(x, z)\ u(z) - \int_{\ssdboundary_N(x,r)}\!\!\!\!\!\!\!\!\!\!\!\!\green^{\ball}(x, z)\ h(z) \ud z
        }{%
            \int_{\ssd(x, r)}\!\!\!\!\!\!\!\!\!\!\!\!\green^{\ball}(x, y)\ f(y)\ud y
        }\label{eq:BoundaryIntegralEquationSimonov}
\end{align}
As with the BIE for WoS (\eqref{WalkOnSpheresIntegral}), the solution value $u(z)$ is the only unknown in this equation: at points $z \in \ssdboundary_N$ the normal derivative $\nicefrac{\partial u}{\partial n}$ is given by the fixed Neumann data along \(\partial \domain\), and is not needed at points $z \in \ssdboundary_{\ball}$, where $G^B(x, z) = 0$. Since only one quantity is unknown, estimators for this equation need not branch.

\citet{Simonov:2008:NeumannWos}, and later \citet{Ermakov:2009:WOH}, take a parallel approach on domains \(\domain\) with \emph{convex} Neumann boundaries \(\partial \domain_N\). In particular, they use regions formed by intersecting $\domain$ with a ball $\ball$ whose radius is the distance to the Dirichlet boundary, $d_\text{Dirichlet} \coloneqq \norm{x - \closest{x}(\partial \domain_D)}$. Hence, $\ball$ can contain a subset of the Neumann boundary $\partial \domain_N$. In the convex case, such regions are automatically star-shaped. To handle arbitrary domains, WoSt instead uses visibility information to obtain star-shaped regions even near nonconvex Neumann boundaries (which in general can yield a radius \(r \le d_\text{Dirichlet}\)), as we will see in \secref{SamplingStarShapedRegions}.

\subsection{The WoSt Estimator}\label{sec:WoStEstimator}
\begin{myTitledBox}{Walk on Stars Estimator}
A recursive single-sample estimator for \eqref{BoundaryIntegralEquationSimonov} is given by
\begin{align}
    \diffeqnSplitEq{,}{%
        \!\widehat{u}(x_k)
    }{%
        \!\frac{\poisson^{\ball}(x_k, x_{k+1})\ \widehat{u}(x_{k+1})}{\alpha(x_k)\ p^{\ssdboundary(x_k,r)}\!(x_{k+1})}\ -\ \!\frac{\green^{\ball}(x_k, z_{k+1})\ h(z_{k+1})}{\alpha(x_k)\ p^{\ssdboundary_N(x_k,r)}\!(z_{k+1})}
    }{%
        \!\frac{\green^{\ball}(x_{k}, y_{k+1})\ f(y_{k+1})}{\alpha(x_k)\ p^{\ssd(x_k,r)}(y_{k+1})}
    }{\coloneqq}\label{eq:BoundaryIntegralEquationEstimator}
\end{align}
where
\begin{itemize}
   \item the points $x_{k+1} \in \ssdboundary$, $z_{k+1} \in \ssdboundary_N$, and $y_{k+1} \in \ssd$ are sampled from the probability densities $p^{\ssdboundary}$ (\secref{RandomWalkStarShapedDomains}), $p^{\ssdboundary_N}$ (\secref{SamplingNeumannBoundaryConditions}), and $p^{\ssd}$ (\secref{SourceSampling}), \resp{}
    \item $r$ is chosen so that $\ssd(x_k,r)$ is star-shaped (\secref{RandomWalkStarShapedDomains}).
\end{itemize}
\end{myTitledBox}

At a high level, each step of WoSt accumulates contributions from the Neumann data $h$ and source term $f$. For mixed Dirichlet-Neumann problems, the walk terminates in $\partial \domain^{\varepsilon}_D$, using the Dirichlet data \(g\) as the solution estimate, \ie, $\widehat{u}(x_k) \coloneqq g(\closest{x}_k(\partial \domain_D))$. For pure Dirichlet problems, WoSt reduces to WoS; for pure Neumann problems we apply Tikhonov regularization (\secref{WoSEstimator}). We first discuss how to sample the next step $x_{k+1}$ (\secref{RandomWalkStarShapedDomains}), followed by sampling procedures for \(h\) and \(f\) (Sections \ref{sec:SamplingNeumannBoundaryConditions}, \ref{sec:SourceSampling}).

\subsection{Random Walk on Star-Shaped Regions}
\label{sec:RandomWalkStarShapedDomains}

The next walk location is importance sampled from the Poisson kernel for a ball centered at the current point $x_k$, \ie, $x_{k+1} \sim p^{\ssdboundary} = \poisson^{\ball}(x_k, x_{k+1})$. For a Poisson equation in \(\R^3\), this kernel is given by
\begin{equation}
    \label{eq:SolidAngleMeasure}
    \poisson^{\ball}_{\text{3D}}(x_k, x_{k+1}) = \frac{n_{x_{k+1}} \cdot (x_{k+1} - x_k)}{4\pi \norm{x_{k+1} - x_k}^3}.
\end{equation}
We can use the same sampling density for a screened Poisson equation, since its corresponding kernel simply multiplies $\poisson^{\ball}$ by a constant in $[0, 1)$ determined by the absorption coefficient (\eqref{ScreenedPoissonKernelsSphere}).

\eqref{SolidAngleMeasure} coincides with the \emph{signed solid angle} subtended by $\ssdboundary$ at $x_{k+1}$ with respect to $x_k$ \citep{Barill:2018:FWN,Feng:2023:WND}. In rendering, this term appears in the \emph{light transport equation (LTE)} \citep[Equation 14.15]{Pharr:2016:PBR}. Unlike the BIE, the LTE multiplies $P^B$ with a binary \emph{visibility} $V(x,y)$ that equals 1 if $x$ and $y$ are mutually visible. Visibility ensures the product $V(x_k, y)\ \poisson^{\ball(x_k, r)}(x_k, y)$ is nonnegative: positive if $y$ is visible from $x_k$; zero otherwise. Through a change of variables, this product can be importance sampled via \emph{directional sampling}, \ie, cast a ray from $x_k$ in a direction $\direction \sim \smash{p^{\Sph^{N-1}}(\direction)} = \nicefrac{1}{|\Sph^{N-1}|}$ uniformly sampled from the unit sphere, and find its first intersection with $\ssdboundary$:
\begin{equation}
    \label{eq:DirectionalSampling}
    x_{k+1} \coloneqq x_k + t_\partial \direction,\ t_\partial \coloneqq \min\!\curly{t\!\in\![0, +\infty): x_k\! +\! t \direction\! \in\! \ssdboundary(x_k, r)}.
\end{equation}
We refer to \citet{veach1995optimally} for details on the relationship between area sampling and directional sampling.%
\footnote{If $x_k$ lies on the boundary, then the ray origin should be offset slightly along the inward boundary normal to avoid self-intersections; see \citet{Wachter:2019:SelfIntersection}.}

\begin{figure}[t]
    \centering
    \includegraphics{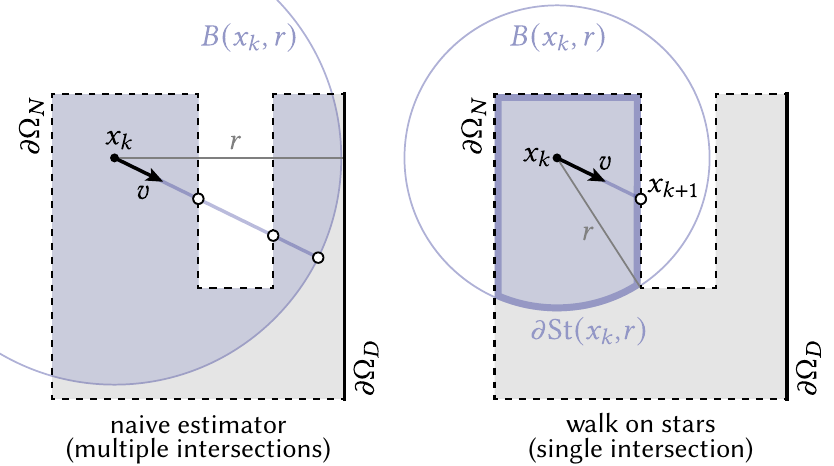}
    \caption{\emph{Left:} For a ball $\ball(x_k, r)$ whose radius $r$ is the distance to the Dirichlet boundary $\partial \domain_D$, the solution to a Poisson equation has to be estimated at all ray intersections, sampled proportional to signed solid angle. \emph{Right:} WoSt instead restricts $\ball \cap \Omega$ to be \emph{star-shaped} relative to $x_k$ to avoid more than one intersection, estimating the PDE solution at the first intersection point $x_{k+1}$ on either $\partial \ball$ inside $\domain$, or the Neumann boundary $\partial \domain_N$ inside $\ball$.}
    \label{fig:MultipleIntersectionsVsStarSampling}
\end{figure}

\subsubsection{Non-Visible Regions}\label{sec:BranchingEstimator}
Since Brownian motion can effectively ``walk around corners'' (\figref{BMvsRBM}), the BIE has no visibility term.  Hence, the solution value at $x_k$ can depend on non-visible points \(x_{k+1}\), complicating use of directional sampling.  In particular, if the subdomain around \(x_k\) is nonconvex, then a na\"{i}ve strategy is to estimate \(u_{x+1}\) at \emph{all} intersections along a ray from \(x_k\) (\figref{MultipleIntersectionsVsStarSampling}, \emph{left}), yielding a branching walk that increases in size exponentially. One could instead use just a single randomly selected intersection, but this approach yields extremely high variance (see \figref{MultipleIntersections}), for two reasons. First, the recursive solution estimate must be multiplied by the number of intersections to account for the expected contribution from each intersection, causing a blowup in value as walk length increases.  Second, the Poisson kernel alternates sign along consecutive intersection points along a ray, yielding unstable estimates due to cancellation~\citep[Chapter 4]{kalos2009monte}.%
\footnote{One might overcome these issues by decomposing the BIE into independent integrals---each with a Poisson kernel that is entirely positive or negative---then randomly select one integral for sampling in proportion to the area over which it is integrated. Unfortunately, it is unclear how to efficiently perform such a decomposition.}
%
These issues are also the root cause of high variance and bias in the walk on boundary method \citep{Sabelfeld:2013:RWB, Sugimoto:2023:WoB}. Moreover, using just the first intersection leads to a biased estimator, as we explain in \appref{FirstIntersection}.


\begin{figure}[t]
    \centering
    \includegraphics{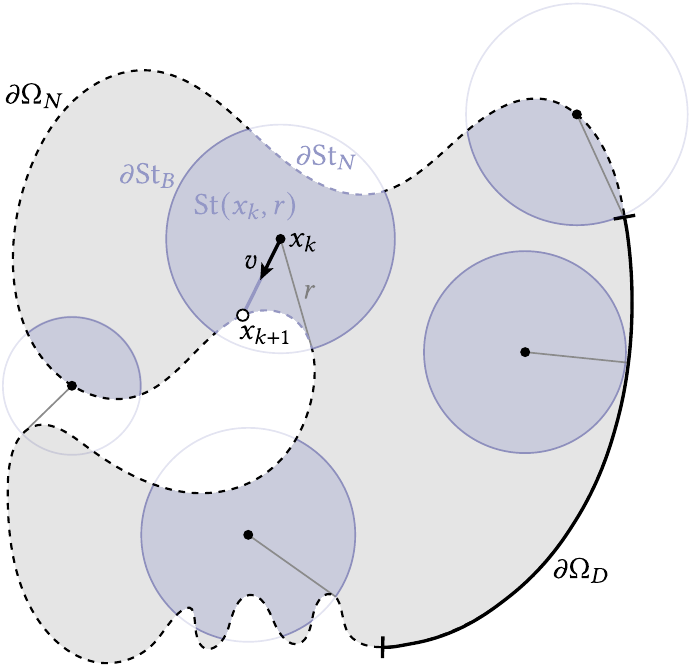}
    \caption{We use star-shaped regions defined by intersecting a ball $\ball(x_k, r)$ with the domain $\domain$ and taking the component connected to \(x_k\). The ball does not contain the Dirichlet boundary $\partial \domain_D$: only visible parts of the Neumann boundary $\partial \domain_N$ and the spherical part of $\partial \ball$.}
    \label{fig:WoStStarShapedDomains}
\end{figure}

\subsubsection{Sampling Star-Shaped Regions}
\label{sec:SamplingStarShapedRegions}

To avoid these issues, some past work assumes the \emph{entire} Neumann boundary $\partial \domain_N$ is convex \citep{Simonov:2008:NeumannWos,Ermakov:2009:WOH}, yielding only one intersection for any subdomain $\ball(x_k, r) \cap \domain$ where $r \coloneqq d_\text{Dirichlet}$.  This assumption of course limits the applicability of such estimators.

We instead let $r$ be the minimum of the distance $d_\text{Dirichlet}$ to the Dirichlet boundary, and the distance $d_\text{silhouette}$ to the closest point on the \emph{visibility silhouette} of $\partial \domain_N$ (\secref{GeometricQueries}).  The connected component of $\ball(x_k,r) \cap \domain$ containing \(x_k\) then defines a star-shaped region $\ssd(x_k, r)$.  \figref{WoStStarShapedDomains} shows several examples.  We can thus sample points on the region boundary \(\ssdboundary\) by simply taking the first point along a ray from \(x_k\) that intersects either \(\partial \ball(x_k,r)\) or \(\partial \domain\).

The use of star-shaped subdomains suggests the name \emph{walk on stars}, in analogy with walk on spheres.  Like the original WoS algorithm (and unlike the reflections in \figref{WoSReflections}) WoSt can take large steps when far from the Dirichlet boundary, using small steps mainly near termination.  Though other star-shaped sets could be also used, our approach is motivated by the fact that the closest silhouette point is quite easy to compute---as will be discussed in \secref{GeometricQueries}.

\begin{figure}[b]
    \centering
    \includegraphics[width=\columnwidth]{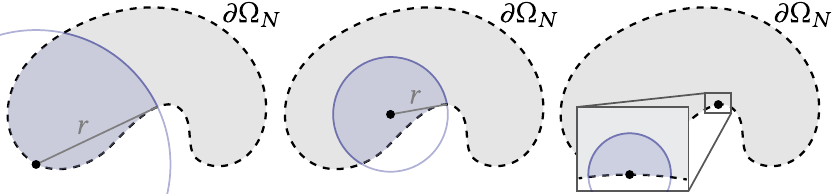}
    \caption{The distance \(r\) to the visibility silhouette shrinks as a query point approaches a concave region on $\partial \domain_N$.}
    \label{fig:ClosestSilhouetteDistanceShrink}
\end{figure}

\subsubsection{Minimum Radius for Star-Shaped Regions}\label{sec:MinimumRadiusParameter}

Near concave parts of the Neumann boundary, the distance to the closest silhouette point on $\partial \domain_N$ shrinks to zero (\figref{ClosestSilhouetteDistanceShrink}), stalling the progress of random walks. We hence limit the radius \(r\) used to define \(\ssd(x_k,r)\) to a user-defined value $r^{\min}$, but still use only the first ray intersection to sample the next point $x_{k+1}$.  This scheme incurs a small amount of bias when $\ssd$ is not star-shaped, since we effectively assume the solution \(u\) is zero on any piece of $\ssdboundary$ not visible from $x_k$ (\figref{MinimumRadiusParameter}, \emph{left})---\appref{FirstIntersection} provides further discussion. As with the parameter $\varepsilon$ for the Dirichlet \(\varepsilon\)-shell $\smash{\partial \domain_D^{\varepsilon}}$, a smaller $r^{{\min}}$ value reduces bias near concave regions of $\partial \domain_N$ at the expense of performance. We study this performance-bias tradeoff in \secref{StoppingTolerances}.  In practice our star-shaped regions tend to be much larger than \(r^{\min}\), even slightly away from a concave boundary.

\newpage

\begin{algo}{\Proc{WalkOnStars}$(x,\ n_x,\ \texttt{onNeumann},\ \varepsilon,\ r^{\min})$}
\label{alg:WalkOnStars}
\begin{algorithmic}[1]
\algblockdefx[Name]{Class}{EndClass}
    [1][Unknown]{\textbf{class} #1}
    {}
\algtext*{EndClass}
\algblockdefx[Name]{FORDO}{ENDFORDO}
    [1][Unknown]{\textbf{for} #1 \textbf{do}}
    {}
\algtext*{ENDFORDO}
\algblockdefx[Name]{IF}{ENDIF}
    [1][Unknown]{\textbf{if} #1 \textbf{then}}
    {}
\algtext*{ENDIF}
\algblockdefx[Name]{IFTHEN}{ENDIFTHEN}
    [2][Unknown]{\textbf{if} #1 \textbf{then} #2}
    {}
\algtext*{ENDIFTHEN}
\algblockdefx[Name]{RETURN}{ENDRETURN}
    [1][Unknown]{\textbf{return} #1}
    {}
\algtext*{ENDRETURN}
\algblockdefx[Name]{COMMENT}{ENDCOMMENT}
    [1][Unknown]{\textcolor{commentpaleblue}{\(\triangleright\) #1}}
    {}
\algtext*{ENDCOMMENT}

\Require A point $x$, the normal $n_x$ at $x$ (undefined if $x \notin \partial \domain_N$), a flag $\texttt{onNeumann}$ indicating whether $x \in \partial \domain_N$, a termination parameter $\varepsilon$, and a minimum radius $r^{\min}$.
   \Ensure A single-sample WoSt estimate $\widehat{u}(x)$ for \eqref{PoissonEquation}.
    \COMMENT[Compute distance to $\partial \domain_D$, or $\infty$ if $\partial \domain_D = \emptyset$ \Comment{\secref{GeometricQueries}}]\ENDCOMMENT
    \State $d_{\text{Dirichlet}},\ \closest{x} \gets \Proc{DistanceDirichlet}(x)$

    \COMMENT[Return boundary value $g$ at $\closest{x}$ if $x \in \partial \domain_D^{\varepsilon}$]\ENDCOMMENT
    \IFTHEN[$d_{\text{Dirichlet}} < \varepsilon$]{\textbf{return} $g(\closest{x})$}
    \ENDIFTHEN

    \COMMENT[Compute distance to the visibility silhouette of $\partial \domain_N$ (Alg. \ref{alg:SilhouetteDistanceNeumann})]\ENDCOMMENT
    \State $d_{\text{silhouette}} \gets \Proc{SilhouetteDistanceNeumann}(x,\ d_{\text{Dirichlet}})$

    \COMMENT[Compute radius for star-shaped region $\ssd(x)$]\ENDCOMMENT
    \State $r \gets \max(r^{\min},\min(d_{\text{Dirichlet}},\ d_{\text{silhouette}}))$

    \COMMENT[Uniformly sample a direction $\direction$ on the unit sphere]\ENDCOMMENT
    \State $\direction \gets \Proc{SampleUnitSphere()}$

    \COMMENT[If $x\!\in\!\partial \domain_N$, sample $\direction$ from hemisphere w/ axis $-n_x$ (Sec. \ref{sec:HemisphericalDirectionSampling})]\ENDCOMMENT
    \IFTHEN[$\texttt{onNeumann}\ \textbf{and}\ n_x \cdot \direction > 0$]{$\direction \gets -\direction$}
    \ENDIFTHEN

    \COMMENT[Intersect ray $x + r \direction$ w/ Neumann part $\ssdboundary_N$, and get first hit]\ENDCOMMENT
    \State $\texttt{didIntersectNeumann},\ p,\ n_p \gets \Proc{IntersectNeumann}(x,\ \direction,\ r)$

    \COMMENT[If there is no hit, intersect spherical part $\ssdboundary_{\ball}$ instead]\ENDCOMMENT
    \IFTHEN[$\textbf{not}\ \texttt{didIntersectNeumann}$]{$p \gets x\ +\ r\ \direction$}
    \ENDIFTHEN

    \COMMENT[Compute single-sample Neumann contribution]\ENDCOMMENT
    \State $z,\ n_z,\ \texttt{pdf}_z \gets \Proc{NeumannBoundarySample}(x,\ r)$\Comment{Alg. \ref{alg:NeumannBoundarySample}}
    \State $\texttt{isValid} \gets \texttt{pdf}_z > 0\ \textbf{and}\ \norm{z - x} < r\ \textbf{and}$ \Comment{\secref{SamplingNeumannBoundaryConditions}}\\ $\hskip\algorithmicindent\hspace{14.25mm} \textbf{not}\ \Proc{IntersectNeumann}(x,\ z - x,\ 1 - 1\text{e-}6)$
    \State $\alpha \gets \texttt{onNeumann}\ ?\ 1/2\ :\ 1$
    \State $\widehat{N} \gets \texttt{isValid}\ ?\ \green^{\ball(x,r)}(x,\ z)\ h(z)\ /\ \alpha\ \texttt{pdf}_z : 0$

    \COMMENT[Compute single-sample source contribution]\ENDCOMMENT
    \State $t_{\text{source}} \sim \green^{\ball(x,r)}(x,\ y)/|\green^{\ball(x,r)}(x)|$\Comment{Sec. \ref{sec:NeumannBoundarySample}}
    \State $y \gets x\ +\ t_{\text{source}}\ \direction$\Comment{Reuse $\direction$ for source sample}
    \State $\widehat{S} \gets \norm{y\!-\!x}\!<\!\norm{p\!-\!x}\ ?\ |\green^{\ball}(x,r)|\ f(y) : 0$\Comment{$f = 0$ if $y \notin \ssd(x)$}

    \COMMENT[Update walk position and normal]\ENDCOMMENT
    \State $x, n_x \gets p, n_p$ \Comment{$n_p$ is undefined if $x \notin \partial \domain_N$}
    \State $\texttt{onNeumann} \gets \texttt{didIntersectNeumann}$
    \RETURN[$\Proc{WalkOnStars}(x, n_x, \texttt{onNeumann}, \varepsilon, r^{\min}) - \widehat{N} + \widehat{S}$]
    \ENDRETURN
\end{algorithmic}
\end{algo}

\newcommand{\SphericalSamplingProblem}{%
\setlength{\columnsep}{1em}
\setlength{\intextsep}{0em}
\begin{wrapfigure}[7]{r}{112pt}
    \centering
    \includegraphics{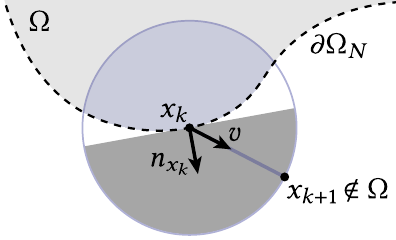}
\end{wrapfigure}
}

\SphericalSamplingProblem{}\subsubsection{Hemispherical Sampling on the Neumann Boundary}\label{sec:HemisphericalDirectionSampling}
When $x_k$ lies on $\partial \domain_N$, sampling \(v\) from the unit sphere can yield points $x_{k+1}$ outside $\domain$ (see inset). Here we instead sample \(v\) from the hemisphere around $n_{x_k}$.  This scheme effectively invokes the \emph{reflection principle} of Brownian motion, across the halfplane at the base of the hemisphere \citep{Jacobs:2010:Stochastic}.  A useful consequence is that the $\alpha(x_k) = 1/2$ in the denominator of the first term in \eqref{BoundaryIntegralEquationEstimator} is canceled by the factor \(1/2\) we get from sampling a hemisphere rather than a sphere, preventing our recursive estimator from picking up a multiplicative factor of two each time a walk reaches $\partial \domain_N$.  Note that if \(\partial \domain_N\) is concave at \(x_k\), we again incur a small amount of bias (\figref{MinimumRadiusParameter}, \emph{right}).

\subsection{Sampling Neumann Boundary Conditions}\label{sec:SamplingNeumannBoundaryConditions}

For problems with nonzero Neumann conditions, we must evaluate the second term in \eqref{BoundaryIntegralEquationEstimator}.  To do so, we sample a point $z_{k+1}$ uniformly on the Neumann boundary $\partial \domain_N$, adding a contribution $h(z_{k+1})$ only if $z_{k+1}$ is also contained in $\ssdboundary_N$.  (\appref{SamplingNeumannBoundaryConditionsBias} explains why this term is not estimated using the next location $x_{k+1}$.)  This estimate remains unbiased, since we effectively integrate the same function (\(h\) restricted to \(\ssdboundary_N\)) over a larger domain.  Sampling the entire Neumann boundary leads to high variance in the estimator, as most samples will not lie on $\ssdboundary_N$.  Likewise, rejection sampling is prohibitively expensive since \(\ssdboundary_N\) can be \emph{much} smaller than \(\partial \domain_N\). In \secref{NeumannBoundarySample}, we hence describe a strategy for efficiently generating samples $z_{k+1}$ close to $x_k$, which significantly reduces variance.

\newcommand{\SourceSamplingFigure}{%
\setlength{\columnsep}{1em}
\setlength{\intextsep}{0em}
\begin{wrapfigure}[8]{r}{90pt}
    \centering
    \includegraphics{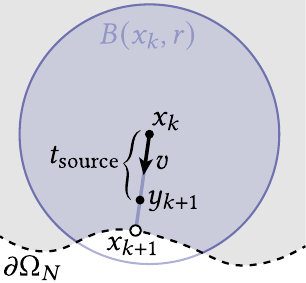}
\end{wrapfigure}
}

\subsection{Sampling the Source Term}\label{sec:SourceSampling}
Finally, we sample a point $y_{k+1} \in \ball(x_k, r)$ to estimate the interior integral in \eqref{BoundaryIntegralEquationEstimator} (\algref{WalkOnStars}, \emph{lines} 24-27). We reuse the\SourceSamplingFigure{}ray direction $\direction$ we sampled to generate $x_{k+1}$ (\eqref{DirectionalSampling}), and set $y_{k+1} \coloneqq x_k + t_\text{source} \direction$, where we sample the distance $t_\text{source} \sim p(t_\text{source}) \propto \green^{\ball}(x_k, x_k + t_\text{source} \direction)$. If the sampled distance $t_\text{source}$ is greater than the distance $t_\partial \coloneqq \norm{x_{k+1}-x_k}$, then the point $y_{k+1}$ is outside the star-shaped region $\ssd(x_k,r)$, and we reject it (see inset). As in \secref{HemisphericalDirectionSampling}, (re)using a hemispherical direction cancels $\alpha(x_k) = 1/2$ when $x_k \in \partial \Omega_N$.

\begin{figure}[t]
    \centering
    \includegraphics{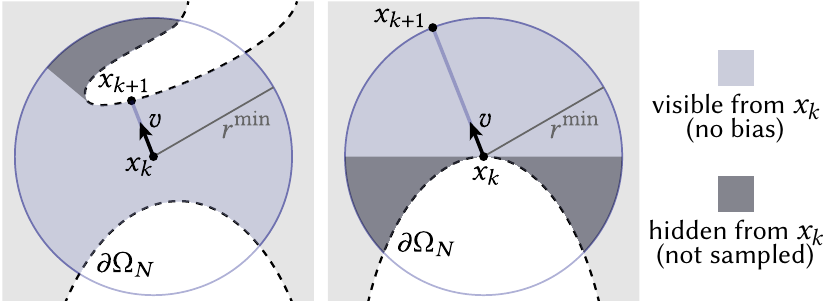}
    \caption{WoSt uses balls with radius no smaller than $r^{\min}$ to prevent walks from stopping near concave Neumann boundaries. \emph{Left:} We only sample parts of $\partial \domain_N$ directly visible to $x_k$ inside any ball $\ball(x_k, r^{\min})$, implicitly assuming the function $u$ is zero elsewhere. \emph{Right:} Hemispherical boundary sampling ensures the next walk location $x_{k+1}$ does not leave the domain, but incurs a small bias in $\widehat{u}$ when $x_k$ lies on a concave boundary.}
    \label{fig:MinimumRadiusParameter}
\end{figure}

\subsection{Final Estimator}\label{sec:WalkOnStarsFinalEstimator}
Our final WoSt estimator is defined recursively as:
\begin{equation}
    \label{FinalWalkOnSpheresEstimator}
    \widehat{u}(x_k) \coloneqq
    \begin{cases}
        g(\overline{x}_k), \!\!& \overline{x}_k \in \partial\Omega_D^{\varepsilon}, \\
        \widehat{u}(x_{k + 1}) - \widehat{N} + \widehat{S} \!\!& \text{otherwise},
    \end{cases}
\end{equation}
where the next walk location $x_{k+1}$ in $\Omega$ or on $\partial \Omega_N$ is sampled using the procedure in \secref{RandomWalkStarShapedDomains}, and the non-recursive Neumann and source contributions $\widehat{N}$ and $\widehat{S}$ are provided in \algref{WalkOnStars}, lines 23 and 27, \resp{}\ This estimator maintains the general structure of a WoS estimator, and thus introduces little implementation overhead.

\section{Geometric Queries}
\label{sec:GeometricQueries}

In general, WoSt works with any boundary representation that supports the following queries:
\begin{enumerate}[label=\textbf{Q.\arabic*}]
    \item \label{enum:cpq} closest point queries to $\partial \domain_D$,
    \item \label{enum:csq} closest silhouette point queries to $\partial \domain_N$,
    \item \label{enum:riq} ray intersection queries against $\partial \domain_N$, and
    \item \label{enum:psq} point sampling queries on $\partial \domain_N$.
\end{enumerate}
Since none of these queries require the boundary to have a well-defined inside/outside, it need not be watertight, and can have cracks, holes, or self-intersections---see in particular \appref{WalkOnStarsDoubleSided} for a discussion of open domains and double-sided boundary conditions.

In principle, queries \ref{enum:cpq}--\ref{enum:psq} could be evaluated for, say, spline patches or implicit surfaces (\secref{LimitationsAndFutureWork}); we focus exclusively on triangle meshes.  In particular, both \ref{enum:cpq} and \ref{enum:riq} use standard closest point and ray intersection algorithms (not detailed here); \secrefs{SNCH,NeumannBoundarySample} describe closest silhouette point queries (\ref{enum:csq}) and point sampling queries (\ref{enum:psq}), \resp{}\  All queries use a \emph{bounding volume hierarchy} (BVH) to achieve amortized sublinear scaling in the number of triangles; our basic approach is to add normal information to the BVH already used by WoS \citep[Section 5.1]{Sawhney:2020:MCGP}.  In particular, we build a standard BVH to perform closest point queries to the Dirichlet boundary (\ref{enum:cpq}), and a separate SNCH (\secref{SNCH}) for all queries of the Neumann boundary (\ref{enum:csq}-\ref{enum:psq}), using normal information only for \ref{enum:csq}.  In practice, all queries needed to implement WoSt on triangle meshes are supported by the \texttt{FCPW} library of \citet{Sawhney:2021:FCPW}; see also \appref{Pseudocode} for pseudocode.

\subsection{Closest Silhouette Point Queries}
\label{sec:SNCH}

The silhouette of a triangle mesh relative to a given direction \(v\) occurs along a set of edges \(e\) that satisfy a local silhouette condition.  In particular, \(e\) is a silhouette edge if for each distinct pair of triangles containing \(e\),
\begin{equation}
    \label{eq:SilhouetteDefinition}
    (v \cdot n_1) \cdot (v \cdot n_2) \leq 0,
\end{equation}
\setlength{\columnsep}{1em}
\setlength{\intextsep}{0em}
\begin{wrapfigure}{r}{94pt}
    \centering
    \includegraphics{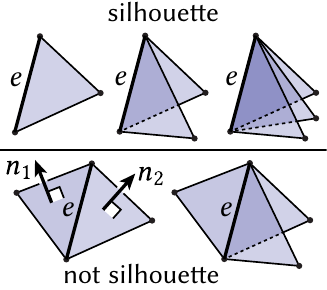}
\end{wrapfigure}
where \(n_1,n_2\) are consistently oriented normals (see inset).  Note in particular that every boundary edge is a silhouette edge.  In WoSt, $v$ is the direction from the current walk location $x$ to its closest point on $e$.  A na\"{i}ve strategy for finding the closest silhouette edge is to use a BVH to locate the closest point to $x$ on all edges, skipping edges not contained in the silhouette.  However, this strategy is highly inefficient when BVH nodes contain large, finely-tessellated regions that are all front- or back-facing (\figref{SphereExhaustiveSilhouetteQuery}): here each edge is examined (and rejected) exhaustively, whereas ideally the whole node should simply be culled.

\begin{figure}[t]
    \centering
    \includegraphics[width=\columnwidth]{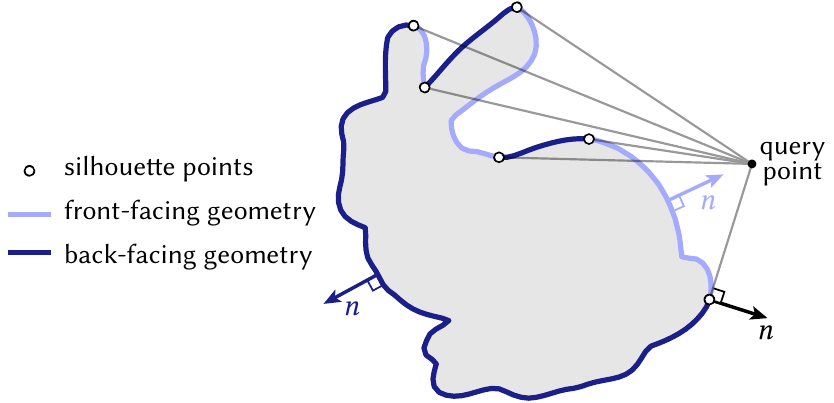}
    \caption{An optimized procedure for finding silhouettes should avoid visiting finely-tessellated geometry that is entirely front- or back-facing relative to the query point.}
    \label{fig:SphereExhaustiveSilhouetteQuery}
\end{figure}

\subsubsection{Spatialized Normal Cone Hierarchy}
\label{sec:SpatializedNormalConeHierarchy}

\begin{figure}[b]
    \centering
    \includegraphics{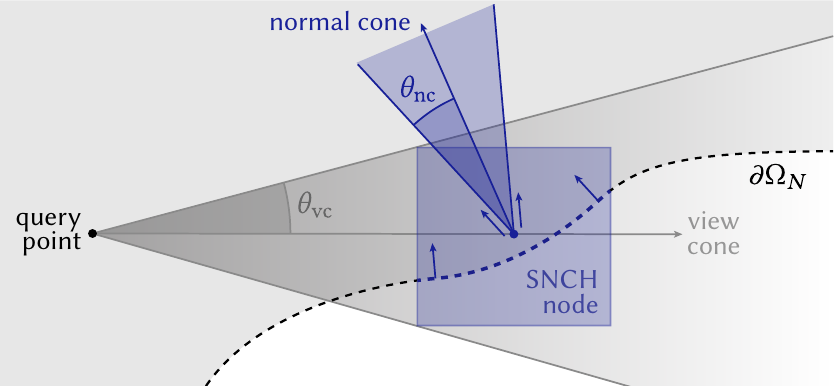}
    \caption{A SNCH tests for a pair of mutually orthogonal directions in a view cone and a node's normal cone to determine whether the node contains a silhouette edge. The geometry inside the node can be skipped if no such pair of directions is found.}
    \label{fig:SpatializedNormalCone}
\end{figure}

To improve scaling, we hence augment our BVH with information about the orientation of the geometry inside each node. In particular, we adopt the \emph{spatialized normal cone hierarchy} (SNCH) of \citet{Johnson:2001:SNCH}. Each node of a SNHC stores not only an axis-aligned bounding box (AABB), but also a \emph{normal cone}. The cone axis is the average normal of all triangles in the node, and the cone half angle \(\theta\) is the maximum angle between the axis and any triangle normal (\figref{SpatializedNormalCone}).  Normal cones can be assembled during BVH construction. We currently use the \emph{surface area heuristic (SAH)} \citep{Wald:2007:SAH}; performance could be further improved via the \emph{surface area orientation heuristic (SAOH)} of \citet[Section 4.4]{Estevez:2018:Importance}, which clusters primitives according to both proximity and alignment.

\subsubsection{Closest Silhouette Point Traversal}
\label{sec:ClosestSilhouettePointTraversal}

To perform a silhouette query, we traverse the SNCH in depth-first order (\algref{SilhouetteDistanceNeumann}).  For each node \node{} in this traversal we build a \emph{view cone} rooted at $x$. The cone's axis points toward the center of \node{}, and its half-angle tightly bounds the AABB (\figref{SpatializedNormalCone}); we then check if the view cone and the node's normal cone contain a pair of mutually orthogonal directions.  If this test fails, all triangles in \node{} must be front- or back-facing relative to the query point, and it can be skipped.  In the context of WoSt, an upper bound on the size of a star-shaped region \(\ssd(x)\) is given by the distance $d_\text{Dirichlet}$ from \(x\) to the Dirichlet boundary (\secref{SamplingStarShapedRegions}). To further improve query efficiency we can hence restrict the search to the radius $r^{\text{max}} = d_{\text{Dirichlet}}$.

\subsection{Point Sampling Queries}
\label{sec:NeumannBoundarySample}

Recall that for problems with nonzero Neumann conditions, we sample points from \(\partial \domain_N\) (\secref{SamplingNeumannBoundaryConditions}).  To increase the likelihood that these points lie on \(\ssdboundary_N(x)\), we adopt a \emph{hierarchical importance sampling} strategy used in rendering to accelerate next-event estimation \citep{Estevez:2018:Importance}. In particular, during each step of BVH traversal we select only a single \emph{random} child whose AABB intersects $\ssd(x)$.  To give preference to nodes closer to the query point $x$, we sample according to the free-space Green's function $\smash{\green^{\R^N}}$ (\algref{NeumannBoundarySample}, \emph{lines} 12-27), rather than the Green's function for a ball (which becomes negative outside \(\ssd(x)\)).  Once we reach a leaf node, we uniformly sample a point from the leaf triangles with respect to surface area (see \algref{SampleTriangleInSphere}, \emph{lines} 2--9 and \algref{NeumannBoundarySample}, \emph{line} 6--10). This point is not guaranteed to lie on $\ssdboundary_N$, but is much more likely to do so compared to uniformly sampling all of $\partial \domain_N$. \citet[Section 5.4]{Estevez:2018:Importance} describe further improvements to this traversal strategy.

\section{Evaluation}\label{sec:Evaluation}
In this section, we discuss practical considerations relating to our WoSt estimator such as stopping tolerances, and use several synthetic tests to evaluate its effectiveness.  We use a multicore CPU-based implementation, and achieve essentially linear scaling; unless otherwise noted, all experiments used a 64-core 3rd Generation Intel Xeon workstation.  As seen in, \eg{}, \figref{Lungs} \emph{(left)} and discussed by \citet[Section 7.5]{Sawhney:2020:MCGP}, the preprocessing cost of building a BVH is typically not significant (on the order of seconds), especially in contrast to finite element mesh generation (on the order of minutes to hours).  Test problems are encoded much like a scene in a photorealistic renderer \cite{Pharr:2016:PBR}, \eg{}, using meshes or other boundary representations to describe \(\partial \domain\), and callback functions to encode the functions \(f\), \(g\), and \(h\) in \eqref{PoissonEquation}.  See \citet[Section 5]{Sawhney:2020:MCGP} for further discussion.

\begin{figure}[t]
    \centering
    \includegraphics{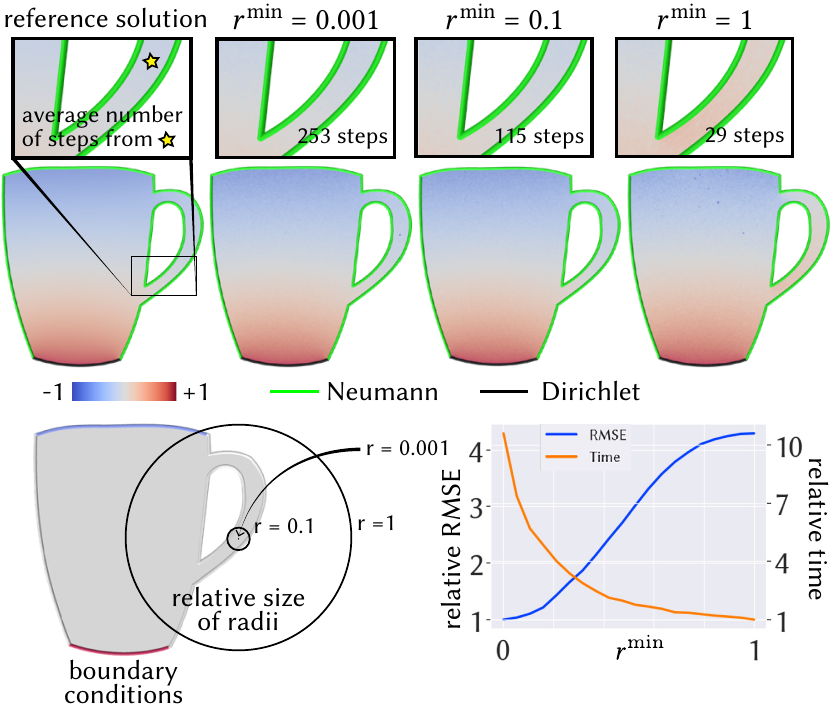}
    \caption{Much like the parameter for the \(\varepsilon\)-shell, WoSt introduces an $r^{{\min}}$ parameter to help walks make progress near concave parts of the Neumann boundary.  Walks generally converge faster with larger values for $r^{{\min}}$, with run-time improvements outweighing the relative increase in bias.}
    \label{fig:MinimumRadius}
\end{figure}

\subsection{Stopping Tolerances}
\label{sec:StoppingTolerances}

Like WoS, WoSt terminates walks in an \(\varepsilon\)-shell $\partial \domain_{D}^{\varepsilon}$ around the Dirichlet boundary, yielding a small, controllable amount of bias.  The performance-bias tradeoff is also relatively insensitive to the parameter \(\varepsilon\).  See \citet{Binder:2012:WosRate} and \citet[Section 6.1]{Sawhney:2020:MCGP} for more detailed analysis and experiments.

The minimum radius parameter \(r^{\min}\) from \secref{MinimumRadiusParameter} also incurs bias near concave parts of $\partial \domain_N$.  \figref{MinimumRadius} examines the effect of this parameter---compared to $\varepsilon$, $r^{{\min}}$ typically exhibits a more sensitive performance-bias tradeoff, but with run-time improvements again outweighing the small increase in bias.  In all other experiments we scale models to fit in a unit sphere, and use \(\varepsilon = r^{\min} = 0.001\).  Adaptively picking $r^{\min}$ based on local boundary curvature may yield even better performance/lower bias.

\begin{figure}[t]
    \centering
    \includegraphics[width=\columnwidth]{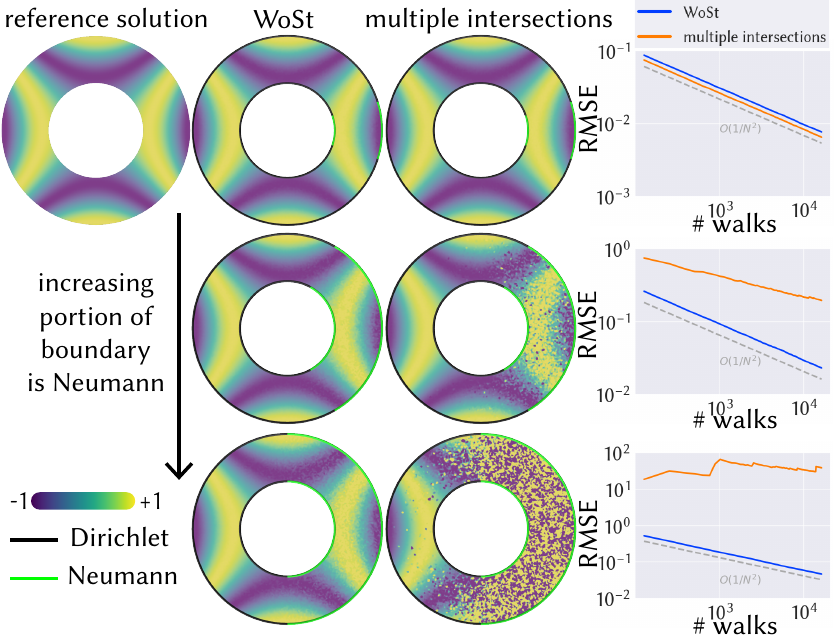}
    \caption{Here we solve for a known reference function, using its normal derivatives to specify Neumann conditions on an increasingly large part of the boundary.  WoSt exhibits the expected Monte Carlo convergence rate, whereas the estimator based on multiple ray intersections from \secref{BranchingEstimator} quickly blows up.}
    \label{fig:MultipleIntersections}
\end{figure}

\begin{figure}
    \centering
    \includegraphics[width=\columnwidth]{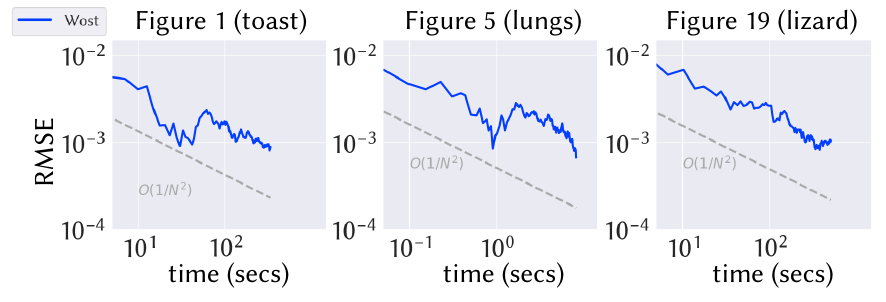}
    \caption{WoSt exhibits the expected rate of convergence for a Monte Carlo estimator, shown here for eight fixed points on each example from \secref{GeometricScaling}. Reference solutions are also computed via WoSt with $2^{16}$ walks per point, as there is no analytical solution and no feasible alternatives to compute it. Timings were taken on an 8 core M1 MacBook Pro.}
    \label{fig:RMSEMainFigures}
\end{figure}

\subsection{Convergence}
\label{sec:Convergence}

As with most Monte Carlo estimators, WoSt exhibits error of magnitude about \(O(1/\sqrt{N})\) with respect to the number of walks \(N\) (\figref{MultipleIntersections}), suggesting that any bias has little impact on overall accuracy.  In general we observe that variance tends to be higher (but still predictable) in regions dominated by Neumann boundaries, due to longer walk lengths.

\begin{figure}[t]
    \centering
    \includegraphics[width=\columnwidth]{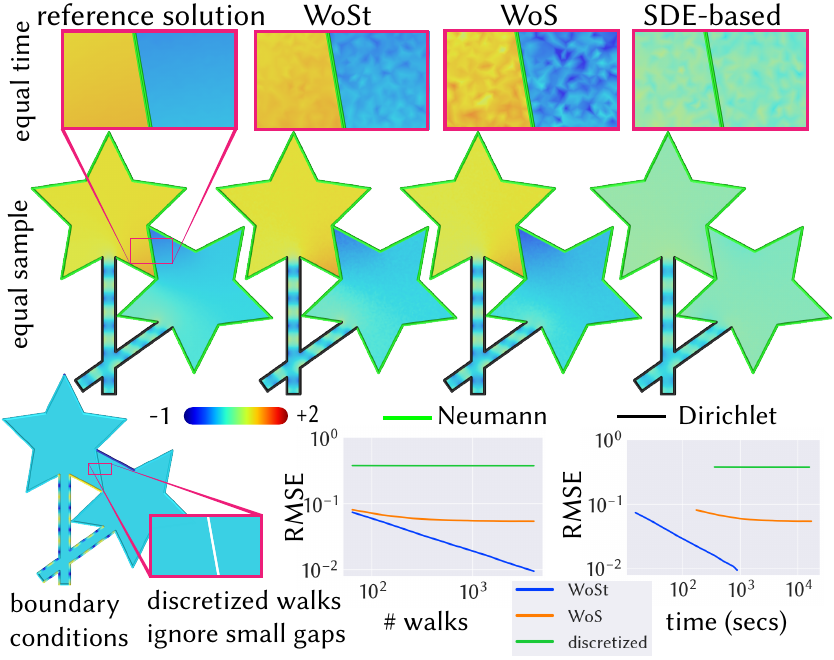}
    \caption{For an equal number of walks, WoSt is significantly more efficient than both WoS with discretized boundary reflections (\secref{WoSEstimator}), as well as SDE-based estimators (\secref{RelatedWork}). Timings were taken on an 8 core M1 MacBook Pro.}
    \label{fig:McComparisons}
\end{figure}

\newcommand{\SDEWalkFigure}{%
\setlength{\columnsep}{2em}
\setlength{\intextsep}{0em}
\begin{wrapfigure}[11]{r}{96pt}
    \centering
    \includegraphics{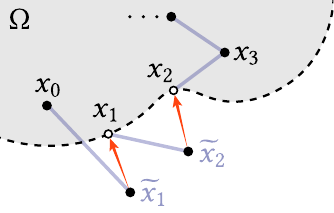}
    \caption{SDE schemes project walks that leave the domain to the boundary, where the walk continues as usual.}
    \label{fig:SDEWalk}
\end{wrapfigure}
}

\SDEWalkFigure{}
\subsection{Comparisons}
\label{sec:MonteCarloComparisons}

WoSt is both significantly faster and less biased than previous Monte Carlo approaches for solving mixed boundary-value problems with comparable parameter settings.  For instance, \figref{McComparisons} compares several methods using comparable parameters: a minimum star radius $r^{\min} = 0.001$ for WoSt, reflection offset $\zeta = 0.01$ for WoS, and step size $l = 0.0001$ for the SDE-based method. As discussed in\secref{WoSEstimator}, WoS with boundary reflections suffers from bias buildup due to long walks that stick to the Neumann boundary. Methods based on reflecting SDEs perform even worse \citep{Constantini:1998:RBM} (see inset), as they incur bias not only on the boundary, but also in the interior.  The na\"{i}ve estimator from \secref{BranchingEstimator}, which selects one intersection at random, also exhibits massive error as we increase the size of the Neumann boundary (\figref{MultipleIntersections}, \figloc{center right}).  In contrast to all these methods, the star-shaped regions used by WoSt enable one to take large steps without incurring significant bias.

\subsection{Pure Neumann Problems}
\label{sec:PureNeumannProblems}

\begin{figure}
    \centering
    \includegraphics[width=\columnwidth]{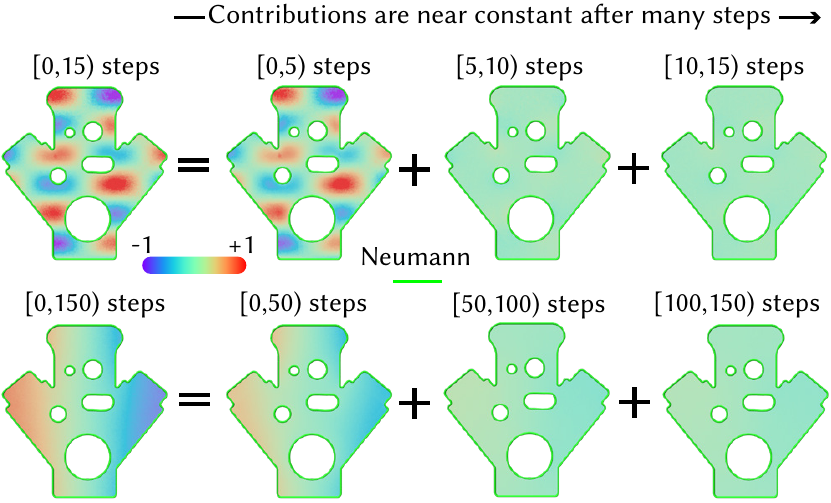}
    \caption{The solution to a Poisson equation with pure Neumann conditions is uniquely defined up to an additive constant. \emph{Top:} Local details in the PDE solution are often resolved by the first few steps of a WoSt random walk, with near-constant contributions from latter steps. \emph{Bottom:} More steps are typically needed to resolve lower frequency global details.}
    \label{fig:PureNeumannConstant}
\end{figure}

\begin{figure}
    \centering
    \includegraphics[width=\columnwidth]{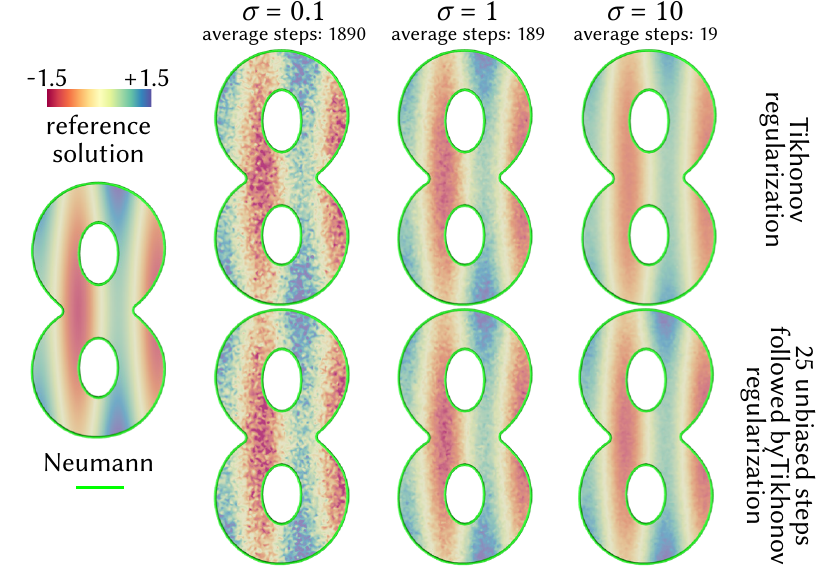}
    \caption{\emph{Top:} For pure Neumann problems, a small Tikhonov parameter \(\sigma\) yields long walks and high variance, while larger \(\sigma\) values produce shorter walks with less noise but more bias. \emph{Bottom:} Since the solution is often well-resolved by short walks (\figref{PureNeumannConstant}), we apply regularization only to walks longer than a given length---yielding both less noise and bias.}
    \label{fig:Tikhonov}
\end{figure}

The solution to a Poisson equation with pure Neumann conditions is determined only up to an additive constant. When we solve such a PDE with WoSt, we observe that high frequency details in the PDE solution are often resolved by the first few steps of a random walk, while the contribution from later steps is closer to constant (\figref{PureNeumannConstant}). As discussed in \secref{WoSEstimator}, we use Tikhonov regularization to more effectively handle such problems---\figref{Tikhonov} shows that this approach provides estimates with less noise and smaller bias even with substantial regularization, while ensuring that walk length is not unbounded.  In general the number of steps needed to resolve the solution is problem-dependent---more steps are typically needed when the solution has low-frequency global features.

\begin{figure*}
   \includegraphics{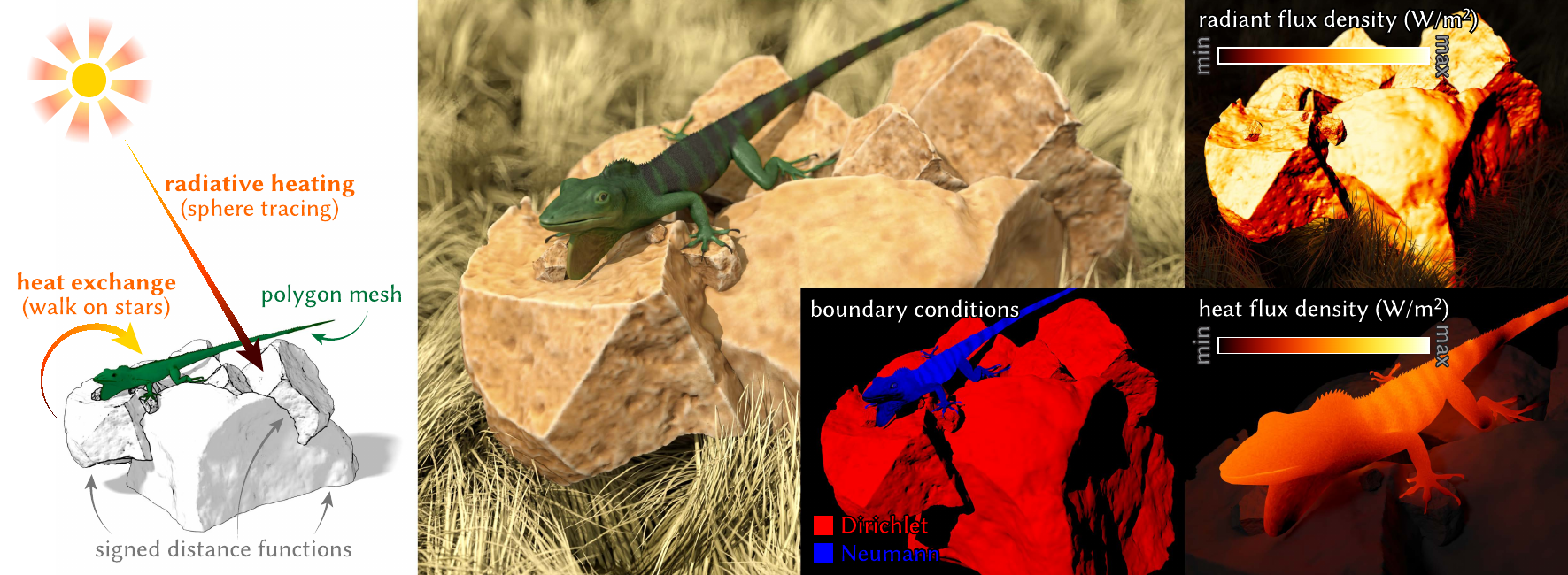}
   \caption{Like Monte Carlo ray tracing, WoSt enables flexible modeling of scenarios that are geometrically and physically complex.  \figloc{Center:} here an ectotherm \emph{(anolis carolinensis)} warms itself on a rock, which in turn is heated by the sun. \figloc{Left:} Like WoS, WoSt can mix and match different geometric representations---here, a polygon mesh and a signed distance function. In this case, it is also easily combined with ray tracing used to determine Dirichlet boundary conditions for the heat transfer problem. \figloc{Right:} Unlike FEM and BEM, WoSt can handle highly-detailed boundary conditions without needing to resolve them on a mesh---here the lizard's texture controls the rate of heat absorption via Neumann boundary conditions, yielding stronger warming along the dark stripes.}
   \label{fig:Lizard}
\end{figure*}

\subsection{Geometric Scaling and Flexibility}
\label{sec:GeometricScaling}

As noted in \secref{Introduction}, a key motivation for developing grid-free Monte Carlo methods is to push simulation methods closer to the geometric complexity seen in photorealistic rendering---and in nature. To stress-test WoSt on complex geometry, we mock up three simulation examples here. Importantly, these examples do not aim to model exact physics or make quantitative predictions---we seek only to examine solver performance in the presence of (i) extremely complex geometry and (ii) large Neumann boundary regions, which are ubiquitous in real physical problems.

\emph{Heat transfer} is a central topic in thermal engineering, with three basic modes: radiation, conduction, and convection. Thermal radiation is well-captured by 1st-order Monte Carlo light transport simulation, whereas conduction and convection involve diffusion, which must be simulated via a 2nd-order method like WoS or WoSt \citep{Bati:2023:CCC}. Thermal convection can also include turbulent advection \ala{} Navier-Stokes, which can be solved via WoS \citep{Rioux-Lavoie:2022:MCFluid} but is not considered here.

Inspired by toast-darkening experiments of \citet{myhrvold2017modernist}, \figref{teaser} models heat transfer from a toaster to a piece of bread, represented by a CT scan with 3.9 million boundary elements. To model diffusive convection, we solve a Laplace equation with large and small Dirichlet values on the heating elements and toaster cavity (\resp{}), and Neumann conditions on the bread.  The solution is evaluated at roughly 2 million boundary points, using 1 walk per point for fast preview and 256 walks for the final solution; on average, WoSt takes \SI{0.166}{milliseconds} per point for each walk (\figref{RMSEMainFigures} plots error versus time). A simple phenomenological model is used to translate surface temperature into color (though more principled models of \emph{Malliard browning} could be used here~\cite{chen2019visual}). We observe a marked difference between the temperature distribution resulting from radiation and convection---emphasizing the necessity of 2nd-order models for accurate thermal predictions.

\figref{Lizard} shows another heat transfer experiment, where Dirichlet conditions induced by solar radiation are used to determine heat absorbed by an ectothermic lizard, modeled via detailed spatially-varying Neumann inflow conditions.  Unlike FEM or BEM, where boundary data must be evaluated ahead of time, Dirichlet data is evaluated \emph{on demand} via sphere tracing \cite{hart1996sphere}.  Scene geometry is represented by a 1.2 million element boundary mesh (for the lizard) and implicit signed distance functions (for the rocks), highlighting the ability of WoSt to work with mixed boundary representations without global meshing.  The solution is evaluated at 285k boundary points using 1024 walks per point, taking on average \SI{0.121}{milliseconds} per point for each walk. As in Monte Carlo rendering (and unlike FEM/BEM), scene setup required no model conversion or meshing---even though data was pulled directly from the internet.

Finally, \figref{Lungs} models oxygen diffusion in the lungs, one of many \emph{Laplacian transport} phenomena with mixed boundary conditions~\citep{Grebenkov:2006:PRBM}.  We make a simplification by using Neumann rather than Robin boundary conditions (which are an important topic for future work---see \secref{LimitationsAndFutureWork}).  While FEM can solve such problems, \emph{meshing} is both a major performance bottleneck and a hindrance for end-to-end robustness.  In this case, even a state-of-the-art method \cite{hu2020fast} yields badly broken geometry; tweaking parameters to capture the correct geometry incurs a full day of compute time, eliminating any advantage of a fast solve. With WoSt we get feedback reliably and immediately in an output-sensitive fashion, here restricted to a cross section. In particular, we evaluate the solution on a $512 \times 512$ grid using 1024 walks per point; WoSt takes on average \SI{0.021}{milliseconds} per point for each walk.

\section{Limitations and Future Work}
\label{sec:LimitationsAndFutureWork}

The principal benefit of WoSt is not simply that it can solve Neumann problems, but rather that it exhibits a speed-bias tradeoff closer to the original WoS algorithm for Dirichlet problems, \ie, it provides large steps and low-variance estimates away from the Dirichlet boundary---especially compared to SDE, WoS, or WoB approaches to the Neumann problem (Sections \ref{sec:RelatedWork} and \ref{sec:MonteCarloComparisons}).  However, Neumann and mixed boundary problems remain fundamentally more challenging than pure Dirichlet problems, with many opportunities for future extension and improvement.

\paragraph{Next-Event, Path-Space, and Bidirectional Estimators}

\begin{figure}
    \centering
    \includegraphics{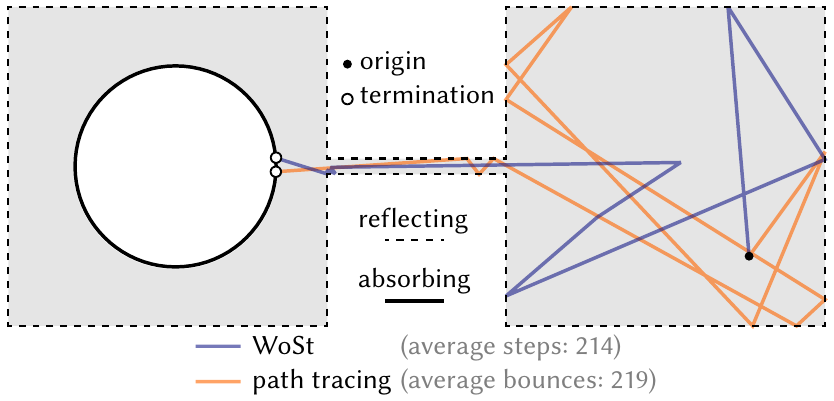}
    \caption{Unidirectional random walk methods such as WoSt and path tracing require many steps or bounces to make their way through a key hole.}
    \label{fig:KeyHole}
\end{figure}

For Neumann-dominated problems, forward random walk estimators must take many steps before obtaining a Dirichlet contribution, resulting in high computation time without a commensurate decrease in variance.  This situation directly parallels forward rendering algorithms that produce long light paths in scenes with predominantly non-absorbing materials.  For instance, \figref{KeyHole} replicates a classic ``keyhole problem'' using both 2D path tracing (assuming perfectly diffuse reflections) and WoSt (with Neumann-dominated boundary). In both cases, average walk length is greater than 200, highlighting a general challenge faced by unidirectional Monte Carlo methods.

In rendering, \emph{next-event estimation} helps by adding a direct illumination contribution at each bounce \citep{Whitted:1980:Improved, Cook:1984:Distributed}.  One could likewise try adding a Dirichlet contribution at each step of WoSt, by allowing subdomains to contain part of the Dirichlet boundary.  Such a scheme would simply need some way to estimate $\nicefrac{\partial u}{\partial n}$ at Dirichlet points (\eqref{BoundaryIntegralEquationSimonov}).

More generally, a \emph{path-space formulation} \citep[Section 14.4.4]{Pharr:2016:PBR} of the BIE may lead to estimators that make more global sampling decisions (\eg{}, via Markov chain Monte Carlo \citep{Veach:1997:Metropolis,kelemen2002simple}) or bidirectional estimators that help connect difficult-to-sample boundary and source data to arbitrary evaluation points \citep{Lafortune:1993:Bidirectional, Veach:1995:Bidirectional}.  The bidirectional WoS method of \citet{Qi:2022:BidirWOS} provides a concrete starting point---along with a rich literature from Monte Carlo rendering \citep{Veach:1997:Metropolis, Vorba:2016:Adjoint, Muller:2017:Practical, Herholz:2019:ZeroVariance, Muller:2019:NIS, Muller:2020:NCV}.

\paragraph{Geometric Queries} The dominant cost in WoSt is evaluating geometric queries (\secref{GeometricQueries}).  While ray- and closest-point queries are already well-optimized \citep{TheEmbreedevelopers:2013:Embree,Wald:2019:Rtx,Krayer:2021:HPD}, closest silhouette point queries could be further accelerated via, \eg{}, better tree construction \citep[Section 4.4]{Estevez:2018:Importance} or intelligent caching of silhouette edges. One might also extend silhouette queries to (neural) implicit surfaces by building on recent range analysis techniques \citep{Sharp:2022:Spelunking}.  Likewise, Neumann point sampling queries could be optimized \ala{} \emph{many-light sampling} \citep[Section 5.4]{Estevez:2018:Importance}, ensuring that Neumann boundary samples lie in star-shaped regions.

\paragraph{Concave Neumann Boundaries} To take larger steps near silhouette points, we could replace the fixed parameter $r^{\min}$ with an adaptive radius based on local curvature estimates \citep{Pottmann:2007:Principal}. Alternatively, one might apply \emph{multi-level Monte Carlo} \citep{Giles:2015:MLMC, Misso:2022:MLMC} to reduce bias by aggregating estimates obtained via progressively smaller values of $r^{\min}$.

\paragraph{Global Information Sharing through Sample Reuse} WoSt estimates solution values independently at each point, often resulting in highly redundant computation.  In concurrent work \citep{Miller:2023:BVC}, we develop a grid-free sample reuse scheme for WoSt inspired by virtual point light (VPL) methods from rendering \citep{Keller:1997:Instant, Dachsbacher:2014:Scalable}.  This method gives unbiased solution estimates at arbitrary points by caching terms of the BIE on the domain boundary, and may open the door to domain decomposition strategies \citep{Chan:1994:Domain} for domains with thin features (\figref{KeyHole}).  Other reuse schemes from rendering such as photon mapping \citep{Hachisuka:2008:Photon, Hachisuka:2009:Photon} and ReSTIR \citep{Bitterli:2020:ReSTIR, Ouyang:2021:ReSTIRGI} surely provide similar opportunities.

\paragraph{Denoising and Geometric Prefiltering}
Since elliptic PDEs have very regular solutions, high-frequency noise in WoSt estimates is nicely mitigated via denoising, as noted by \citet[Figure 13]{Sawhney:2020:MCGP}; here again methods from rendering provide a wealth of opportunities \citep{Zwicker:2015:Adaptive, Chaitanya:2017:Denoising, Schied:2017:FilteringA, Schied:2018:FilteringB, Gharbi:2019:Denoising, Pawel:2019:ReLAX, Nvidia:2022:NRD, Nvidia:2017:Optix}.  Another interesting challenge is how to detect---or even define---silhouettes for geometry with intricate microstructures \citep{Neyret:1998:Modeling}, perhaps through some form of geometric prefiltering \citep{Wu:2019:Prefiltering}.  In particular, the SNCH itself provides a form of prefiltering: nodes higher than the leaves can provide conservative or approximate bounds on \(d_{\text{silhouette}}\) that effectively amount to smoothing out fine-scale geometry (at the cost of bias).  Likewise, SNCH nodes could be built by sampling a distribution (\ala{} microflake models \citep{Heitz:2015:Sggx}) rather than bounding explicit geometry.


\begin{figure}[t]
    \centering
    \includegraphics[width=\columnwidth]{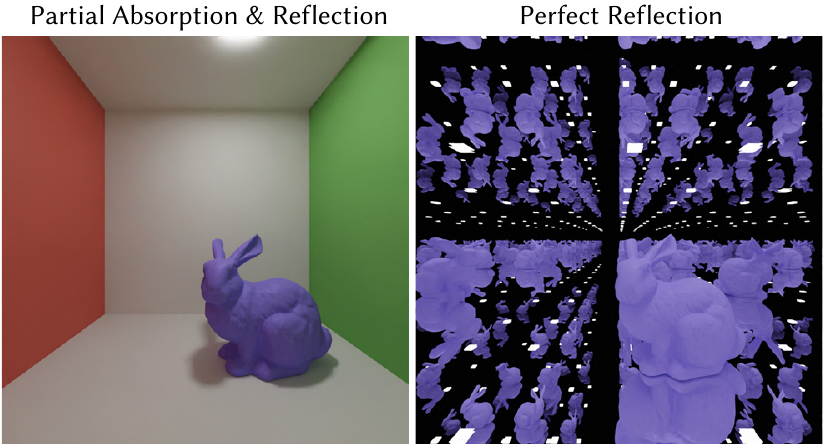}
    \caption{Realistic scenes for visualization and analysis rarely have purely reflecting surfaces. \emph{Left:} A rendered scene with both absorbing and reflecting surfaces. \emph{Right:} A rendered scene of a room full of perfect mirrors.}
    \label{fig:RoomOfMirrors}
\end{figure}

\paragraph{Robin Boundary Conditions}
In rendering, a room full of non-absorbing mirrors yields much longer path lengths than a scene with realistic, partially absorptive materials (\figref{RoomOfMirrors}).  Likewise, perfectly-reflecting Neumann conditions represent an ``extreme case'' for grid-free Monte Carlo methods, since materials in real physical problems are typically reflecting \emph{and} absorbing.  For instance, realistic thermal, electromagnetic, and fluid models often assume \emph{Robin boundary conditions} of the form
\begin{equation}
    \label{eq:RobinBoundaryConditions}
    \alpha u(x) + \beta \frac{\partial u(x)}{\partial n_x} = g(x)\ \text{ on } \partial \domain, \quad \alpha,\beta \in \mathbb{R}
\end{equation}
\citep{Senior:1960:Impedance, Gustafson:1998:Third, Grebenkov:2006:PRBM, Hahn:2012:Heat}.  Incorporating such conditions into WoSt would result in shorter walks and greater efficiency, since (as in rendering) low-throughput walks could be terminated early via Russian roulette \citep[Section 13.7]{Pharr:2016:PBR}.  Support for Robin boundary conditions would enable WoSt to solve PDEs with variable coefficients \citep{Sawhney:2022:VCWoS} and exterior problems  \citep{Nabizadeh:2021:Kelvin} with mixed boundary conditions, since the \emph{Girsanov} and \emph{Kelvin transformations} used by these methods (\resp{}) convert Neumann conditions into Robin conditions.  Work by \citep[Section 1]{Simonov:2017:Robin} provides one possible starting point.

\paragraph{Extension to Other PDEs}

Though we have developed WoSt in the context of (screened) Poisson equations, we believe the basic algorithmic strategy applies more broadly: fundamentally, the WoSt algorithm depends on the structure of the BIE, and BIE formulations are readily available for a variety of other PDEs, including the Helmholtz equation \citep[Chapter 3]{Hunter:2001:BEM}, linear elasticity \citep[Chapter 4]{Hunter:2001:BEM} and the biharmonic equation \citep{Ingham:2012:BIE}. Stochastic integral formulations are also known for Navier-Stokes \citep{Busnello:2005:Probabilistic, Rioux-Lavoie:2022:MCFluid}.

Even within the class of PDEs presented here, geometric scalability will likely pay dividends in well-chosen scientific and engineering contexts---just as it has for Monte Carlo simulation of light transport.  In general, the very different capabilities of grid-free Monte Carlo methods are still largely unexplored in geometric, visual, and scientific computing, with many unique benefits and attractive use cases yet to be discovered.

\begin{acks}
    The authors thank Gautam Iyer for suggesting Tikhonov regularization, and Dario Seyb and Rasmus Tamstorf for fruitful conversations. This work was generously supported by nTopology and Disney Research, NSF awards 1943123, 2212290 and 2008123, Alfred P. Sloan Research Fellowship FG202013153, a Packard Fellowship, NSF Graduate Research Fellowship DGE2140739, and an NVIDIA Graduate Fellowship.
\end{acks}

\bibliographystyle{ACM-Reference-Format}
\bibliography{WalkOnStars}

\newpage

\appendix

\section{Green's Functions and Poisson Kernels}\label{app:GreensFns}
Here we provide expressions for Green's functions and Poisson kernels in free-space and over a ball in 2D and 3D. Our WoSt estimator uses these functions to solve the Poisson and screened Poisson equations via their boundary integral formulation.

\subsection{Poisson Equation}\label{app:GreensFnsPoisson} The free-space Green's functions in 2D and 3D for two points $x$ and $y$ equal:
\begin{equation}
    \label{eq:FreeSpaceGreensFnsPoisson}
    \green^{\R^2}(x,y) = \frac{\log(r)}{2\pi},
    \quad
    \green^{\R^3}(x,y) = \frac{1}{4\pi r},
\end{equation}
where $r \coloneqq \norm{y - x}$. The corresponding Poisson kernels equal:
\begin{equation}
    \label{eq:FreeSpacePoissonKernels}
    \poisson^{\R^2}(x,y) = \frac{n_y \cdot \left(y - x\right)}{2\pi r^2},
    \quad
    \poisson^{\R^3}(x,y) = \frac{n_y \cdot \left(y - x\right)}{4\pi r^3},
\end{equation}
where $n_y$ is the unit normal at $y$.

For a ball $\ball(x, R)$ of radius $R$ centered at $x$, we can derive 2D and 3D Green's functions from the corresponding free-space expressions using the \emph{method of images} \citep{Duffy:2015:Green}. This gives:
\begin{equation}
    \label{eq:GreensFnsPoissonBall}
    \green_{\text{2D}}^{\ball}(x,y) = \frac{\log(R/r)}{2\pi},
    \quad
    \green_{\text{3D}}^{\ball}(x,y) = \frac{1}{4\pi}\left(\frac{1}{r} - \frac{1}{R}\right).
\end{equation}
These functions integrate over the ball $\ball(x, R)$ to:
\begin{align}
    \label{eq:GreensFnsNormPoissonBall}
    |\green_{\text{2D}}^{\ball}(x)| &\coloneqq \int_{\ball(x, R)} \green_{\text{2D}}^{\ball}(x,y) \ud y = \frac{R^2}{4},\\
    |\green_{\text{3D}}^{\ball}(x)| &\coloneqq \int_{\ball(x, R)} \green_{\text{3D}}^{\ball}(x,y) \ud y = \frac{R^2}{6}.
\end{align}
The corresponding Poisson kernels are the same as their free-space counterparts. For any point $y \in \partial \ball$ where $n_y = (y - x)/R$, they simplify to:
\begin{equation}
    \label{eq:PoissonKernelsSphere}
    \poisson_{\text{2D}}^{\ball}(x,y) = \frac{1}{2\pi R},
    \quad
    \poisson_{\text{3D}}^{\ball}(x,y) = \frac{1}{4\pi R^2}.
\end{equation}

\subsection{Screened Poisson Equation}\label{app:GreensFnsScreenedPoisson}

Let $I_n$ and $K_n$ (for $n = 0, 1, \ldots$) denote modified Bessel functions of the first and second kind, \resp{}\ The free-space Green's functions in 2D and 3D for a screened Poisson equation with a positive screening coefficient $\sigma$ then equal:
\begin{equation}
    \label{eq:FreeSpaceGreensFnsScreenedPoisson}
    \green^{\sigma, \R^2}(x,y) = \frac{K_0(r\sqrt{\sigma})}{2\pi},
    \quad
    \green^{\sigma, \R^3}(x,y) = \frac{e^{-r\sqrt{\sigma}}}{4\pi r}.
\end{equation}
The corresponding Poisson kernels are:
\begin{align}
    \label{eq:FreeSpaceScreenedPoissonKernels}
    \poisson^{\sigma, \R^2}(x,y) &= Q^{\sigma, \R^2}(x, y)\ \poisson^{\R^2}(x,y),\\
    \poisson^{\sigma, \R^3}(x,y) &= Q^{\sigma, \R^3}(x, y)\ \poisson^{\R^3}(x,y),
\end{align}
where
\begin{align}
    Q^{\sigma, \R^2}(x, y) &\coloneqq K_1(r\sqrt{\sigma})\ r\sqrt{\sigma},\\
    Q^{\sigma, \R^3}(x, y) &\coloneqq e^{-r\sqrt{\sigma}}\left(r\sqrt{\sigma} + 1\right).
\end{align}
For a ball $\ball(x, R)$, the 2D and 3D Green's functions are:
\begin{align}
    \label{eq:GreensFnsScreenedPoissonBall}
    \green_{\text{2D}}^{\sigma, \ball}(x,y) &= \frac{1}{2\pi}\left(K_0(r\sqrt{\sigma}) - I_0(r\sqrt{\sigma})\frac{K_0(R\sqrt{\sigma})}{I_0(R\sqrt{\sigma})}\right),\\
    \green_{\text{3D}}^{\sigma, \ball}(x,y) &= \frac{1}{4\pi}\left(\frac{\sinh((R - r)\sqrt{\sigma})}{r \sinh(R\sqrt{\sigma})}\right).
\end{align}
Their integral over the ball $\ball(x, R)$ equals:
\begin{align}
    \label{eq:GreensFnsNormScreenedPoissonBall}
    |\green_{\text{2D}}^{\sigma, \ball}(x)| &\coloneqq \int_{\ball(x, R)} \green_{\text{2D}}^{\sigma, \ball}(x,y) \ud y = \frac{1}{\sigma}\left(1 - \frac{1}{I_0(R\sqrt{\sigma})}\right),\\
    |\green_{\text{3D}}^{\sigma, \ball}(x)| &\coloneqq \int_{\ball(x, R)} \green_{\text{3D}}^{\sigma, \ball}(x,y) \ud y = \frac{1}{\sigma}\left(1 - \frac{R\sqrt{\sigma}}{\sinh(R\sqrt{\sigma})}\right).
\end{align}
Finally, the corresponding Poisson kernels at any point $y \in \ball$ equal
\begin{align}
    \label{eq:ScreenedPoissonKernelsSphere}
    \poisson_{\text{2D}}^{\sigma, \ball}(x,y) &= Q^{\sigma, \ball}_{\text{2D}}(x, y)\ \poisson^{\R^2}(x,y),\\
    \poisson_{\text{3D}}^{\sigma, \ball}(x,y) &= Q^{\sigma, \ball}_{\text{3D}}(x, y)\ \poisson^{\R^3}(x,y),
\end{align}
where
\begin{align}
    Q^{\sigma, \ball}_{\text{2D}}(x, y) &\coloneqq \left(K_1(r\sqrt{\sigma}) + I_1(r\sqrt{\sigma})\frac{K_0(R\sqrt{\sigma})}{I_0(R\sqrt{\sigma})}\right)\ r\sqrt{\sigma},\\
    Q^{\sigma, \ball}_{\text{3D}}(x, y) &\coloneqq
        \begin{multlined}[t]
            e^{-r\sqrt{\sigma}}\left(r\sqrt{\sigma} + 1\right)\ + \\
            \left(\cosh(r\sqrt{\sigma})\ r\sqrt{\sigma} - \sinh(r\sqrt{\sigma})\right)\frac{e^{-R\sqrt{\sigma}}}{\sinh(R\sqrt{\sigma})}.
        \end{multlined}
\end{align}
We note that $Q^{\sigma, \ball} \in [0, 1)$ for $\sigma > 0$. Using \eqref{ScreenedPoissonKernelsSphere}, we introduce a multiplicative weight of $Q^{\sigma, \ball}$ in the solution estimate at every step of a WoSt random walk for a screened Poisson equation, since directions are sampled proportionally to $\poisson^{\R^N}$ (\secref{RandomWalkStarShapedDomains}). As we mention in \secref{WoSEstimator}, the cumulative product of $Q^{\sigma, \ball}$ can be used as a Russian roulette probability to terminate walks.

\newcommand{\DoubleSidedFigure}{%
\setlength{\columnsep}{1em}
\setlength{\intextsep}{0em}
\begin{wrapfigure}[9]{r}{88pt}
    \centering
    \includegraphics{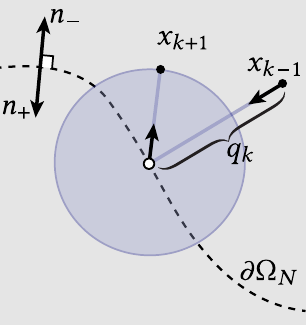}
\end{wrapfigure}
}
\section{Open Domains and Double-Sided Boundaries}
\label{app:WalkOnStarsDoubleSided}

Here we describe how to modify \algref{WalkOnStars} to support double-sided boundary conditions in an open domain $\domain \subset \R^N$. We start with the BIE for double-sided boundary conditions \citep{Costabel:1987:BEM}:
\begin{align}
    \diffeqnSplitDoubleSided{,}{%
        \!\alpha(x)u(x)\!
    }{%
    \!\int_{\partial \domain}\!\!\!\!\!\!\poisson^+\!(x, z)\!\!\left[u^+\!(z)\!-\! u^-\!(z)\right]
    \!-\! \green(x, z)\!\!\left[\frac{\partial u^+\!(z)}{\partial n^+_z}\! -\!\frac{\partial u^-\!(z)}{\partial n^-_z}\right]\!\!\ud z
    }{%
    \!\int_{\domain}\!\!\! \green(x, y)\ f(y) \ud y
    }\label{eq:BoundaryIntegralEquationDoubleSided}
\end{align}
%
where $n^+$ and $n^-$ denote unit outward and inward facing normals on $\partial \domain$, \resp{}, $u^+$ and $u^-$ represent corresponding solution values on either side of $\partial \domain$, and the kernel $P^+(x, z) \coloneqq \nicefrac{\partial G(x, z)}{\partial n^+_z}$.  Since all points are either on the boundary or the domain interior, $\alpha = 1/2$ on $\partial \Omega$ and 1 otherwise.

\DoubleSidedFigure{}The high-level idea is to estimate \eqref{BoundaryIntegralEquationDoubleSided} by choosing an appropriate set of boundary conditions to use in a star-shaped region $\ssd(x_k, r)$ centered at the current walk location $x_k$, \ie, $g^+$ or $g^-$ as the Dirichlet data when $x_k \in \partial \domain_D$, and $h^+$ or $h^-$ as the Neumann data in $\ssd$ when $x_k \in \partial \domain_N$. The choice of boundary conditions depends on whether the boundary $\partial \domain$ is front- or back-facing relative to $x_k$ or $x_{k-1}$ for $k > 0$ (inset). We assume the boundary $\partial \domain$ has a canonical orientation defined by the unit outward normal $n^+_z$ for any point $z \in \partial \domain$, and that the walk's direction of approach towards $\partial \domain$ equals $q_k \coloneqq x_k - x_{k-1}$. For $k = 0$, we require $q_0$ as input to the algorithm to determine the user-specified choice of boundary conditions to use; for instance, setting $q_0 = n^+_{x_0}$ uses $g^+$ if $x_0 \in \partial \domain_D$.

For double-sided Dirichlet boundary conditions, we change line 5 in \algref{WalkOnStars} to return $g^+$ when $q_k \cdot n^+_{\closest{x}_k} > 0$ and $g^-$ otherwise; the dot product determines whether the boundary is back-facing relative to $x_{k-1}$. Before using hemispherical direction sampling on $\partial \domain_N$ to determine the next walk location $x_{k+1}$ (\emph{line} 13), we flip the direction of the boundary normal $n^+_{x_k}$ if $q_k \cdot n^+_{x_k} < 0$; this ensures that $x_{k+1}$ lies on the same side of the boundary as the direction from which the walk approached $\partial \domain_N$. Finally, we let $z_{k+1}$ be a Neumann sample on $\ssdboundary_N$ for double-sided Neumann conditions. In line 23, we then use $h^+$ on $z_{k+1}$ if $(z_{k+1} - x_k) \cdot n^+_{z_{k+1}} > 0$ and $h^-$ otherwise when $x_k \notin \partial \domain_N$; the boundary orientation in this case is determined relative to $x_k$. When $x_k \in \partial \domain_N$, we use $h^+$ as the Neumann data on $z_{k+1}$ if the boundary normal is flipped and $h^-$ otherwise.

There is a non-zero probability for a random walk to wander off to infinity when using WoSt in an open domain or in the exterior of a closed domain; this corresponds to the \emph{non-recurrent} behavior of Brownian motion in 2D and 3D \citep[Chapter 2]{Borodin:2015:Handbook}. \citet{Nabizadeh:2021:Kelvin} use WoS to solve exterior problems with pure Dirichlet boundary conditions outside a closed domain by performing a spherical inversion of the domain. We leave extending their approach to mixed boundary-value problems to future work.

\section{Why Not Always Use the First Intersection}
\label{app:FirstIntersection}

\setlength{\columnsep}{1em}
\setlength{\intextsep}{0em}
\begin{wrapfigure}{r}{90pt}
    \centering
    \includegraphics{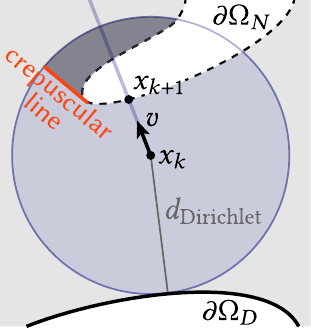}
\end{wrapfigure}
At first glance, the discussion in \secref{BranchingEstimator} prompts a simple idea for generalizing the technique of \citet{Simonov:2008:NeumannWos} and \citet{Ermakov:2009:WOH} to nonconvex domains (which unfortunately does not work): consider the subdomain $\ball(x_k, d_\text{Dirichlet}) \cap \domain$, and sample the next point \(x_{k+1}\) by taking the \emph{first} intersection with $\partial\left(\ball(x_k, d_\text{Dirichlet}) \cap \domain\right)$ along a ray \(v\) from $x_k$, ignoring any later intersections. This is analogous to choosing a BIE subdomain $\genset \subset \ball(x_k, d_\text{Dirichlet}) \cap \domain$ comprising only points in $\ball(x_k, d_\text{Dirichlet}) \cap \domain$ visible to $x_k$.
The boundary of such a subdomain $\genset$ is made up of two components: first, the part of $\partial\left(\ball(x_k, d_\text{Dirichlet}) \cap \domain\right)$ that is visible to $x_k$; second, any crepuscular rays~\citep{gargallo2007minimizing} corresponding to the visibility silhouette on $\partial\left(\ball(x_k, d_\text{Dirichlet}) \cap \domain\right)$ with respect to $x_k$ (see inset).

The second component is the reason why just sampling the first intersection would produce biased results, as one would never sample any points on the crepuscular rays. Even if we considered a modified sampling strategy that allowed for generating points $z$ on these rays, we would additionally need to estimate the normal derivative $\nicefrac{\partial u(z)}{\partial n_z}$ of the solution at $z$ as this data is only known on the Neumann boundary. (An estimate of the solution $u(z)$ is not needed, as the Poisson kernel is zero on crepuscular rays.) By constrast, we shrink the subdomain A to $\ball(x_k, \min\paren{d_\text{Dirichlet}, d_\text{silhouette}}) \cap \domain$ in \secref{SamplingStarShapedRegions} so that it does not contain any crepuscular rays---using just the first intersection then evades the aforementioned problems.

\section{Why Sample the Neumann Term Separately}\label{app:SamplingNeumannBoundaryConditionsBias}
We use two separate samples $x_{k+1}$ and $z_{k+1}$ to estimate the first and second terms in \eqref{BoundaryIntegralEquationEstimator} corresponding to the unknown solution value $u$ and known Neumann data $h$, \resp{}\ However, we could in theory use $x_{k+1}$ to estimate both boundary terms. To do so, we would rewrite the WoSt estimator as follows:
\begin{align}
    \diffeqnSplitEq{,}{%
        \!\widehat{u}(x_k)
    }{%
        \!\frac{\poisson^{\ball}\!(x_k, x_{k+1})\ \widehat{u}(x_{k+1})}{\alpha(x_k)\ p^{\ssdboundary(x_k,r)}\!(x_{k+1})}\ -\ \!\frac{\green^{\ball}\!(x_k, x_{k+1})\ \tilde{h}(x_{k+1})}{\alpha(x_k)\ p^{\ssdboundary(x_k,r)}\!(x_{k+1})}
    }{%
        \!\frac{\green^{\ball}\!(x_{k}, y_{k+1})\ f(y_{k+1})}{\alpha(x_k)\ p^{\ssd(x_k,r)}(y_{k+1})}
    }{\coloneqq}\label{eq:BoundaryIntegralEquationEstimatorAlt}
\end{align}
where
\begin{equation}
    \label{eq:NeumannTilde}
    \tilde{h}(x_{k+1}) \coloneqq
    \begin{cases}
        h(x_{k+1}), \!\!& x_{k+1} \in \ssdboundary_N(x_k,r), \\
        0, \!\!& x_{k+1} \in \ssdboundary_\ball(x_k,r).
    \end{cases}
\end{equation}
In practice, this approach is problematic as the second term in \eqref{BoundaryIntegralEquationEstimatorAlt} is biased: the direction sampling procedure in \secref{RandomWalkStarShapedDomains} never samples a point $x_{k+1}$ on a flat Neumann boundary when $x_k \in \partial \domain_N$, even if $h$ is non-zero there. This is because the Poisson kernel $P^B$ in \eqref{SolidAngleMeasure} (which serves as our sampling density $p^{\ssdboundary}$) is zero for points $x_{k+1}$ where $n_{x_{k+1}} \perp \left(x_{k+1} - x_k\right)$. More generally, even for non-flat boundaries, the ratio $\nicefrac{\green^{\ball}}{p^{\ssdboundary}}$ $\coloneqq \nicefrac{\green^{\ball}}{\poisson^{\ball}}$ in the second term in \eqref{BoundaryIntegralEquationEstimatorAlt} results in high-variance estimates as the Poisson kernel can take on both very large and small values. These issues motivate our use of a separate sample $z_{k+1}$ from the probability density function $p^{\ssdboundary_N(x_k,r)}\!(z_{k+1})$ (\eqref{BoundaryIntegralEquationEstimator}), which we generate using the procedure described in \secref{SamplingNeumannBoundaryConditions}.

\section{Pseudocode}\label{app:Pseudocode}
Here we provide pseudocode for the geometric queries in \secref{GeometricQueries}.

\begin{algo}{\Proc{SilhouetteDistanceNeumann}$(x,\ r^{\texttt{max}}=\infty)$}
\label{alg:SilhouetteDistanceNeumann}
\begin{algorithmic}[1]
\Require A query point $x \in \R^3$ \& a radius $r$ around $x$ to search in.
\Ensure The closest point to $x$ on the visibility silhouette of $\partial \domain_N$.
    \State{\Return $\Proc{DistClosestSilhouette}(\texttt{snch}.\texttt{root},\ x,\ r^{\texttt{max}})$}
\end{algorithmic}
\end{algo}

\begin{algo}{\Proc{DistClosestSilhouette}$(T,x,r,d_T^{{\min}} = 0)$}
\label{alg:DistClosestSilhouette}
\begin{algorithmic}[1]
\algblockdefx[Name]{Class}{EndClass}
    [1][Unknown]{\textbf{class} #1}
    {}
\algtext*{EndClass}
\algblockdefx[Name]{FORDO}{ENDFORDO}
    [1][Unknown]{\textbf{for} #1 \textbf{do}}
    {}
\algtext*{ENDFORDO}
\algblockdefx[Name]{IF}{ENDIF}
    [1][Unknown]{\textbf{if} #1 \textbf{then}}
    {}
\algtext*{ENDIF}
\algblockdefx[Name]{IFTHEN}{ENDIFTHEN}
    [2][Unknown]{\textbf{if} #1 \textbf{then} #2}
    {}
\algtext*{ENDIFTHEN}
\algblockdefx[Name]{RETURN}{ENDRETURN}
    [1][Unknown]{\textbf{return} #1}
    {}
\algtext*{ENDRETURN}
\algblockdefx[Name]{COMMENT}{ENDCOMMENT}
    [1][Unknown]{\textcolor{commentpaleblue}{\(\triangleright\)#1}}
    {}
\algtext*{ENDCOMMENT}
\algblockdefx[Name]{ELSEIF}{ENDELSEIF}
    [1][Unknown]{\textbf{else if} #1 \textbf{then}}
    {}
\algtext*{ENDELSEIF}
\algblockdefx[Name]{ELSE}{ENDELSE}
    {\textbf{else}}
    {}
\algtext*{ENDELSE}

\Require A spatialized normal cone hierarchy $T$, a query point $x \in \R^3$, a radius $r$ around $x$ to search in, and optionally the minimum distance to $T$'s AABB from $x$ (0 if $x$ is inside AABB).
\Ensure Distance from $x$ to closest point on visibility silhouette.
    \IFTHEN[$d_T^{{\min}} > r$]{$\textbf{return}\ r$}\Comment{Ignore nodes outside search radius}
    \ENDIFTHEN
    \IF[$T.\texttt{isLeaf}$]
        \COMMENT[Return distance to closest silhouette edge in leaf node]\ENDCOMMENT
        \FORDO[$e\ \textbf{in}\ T.\texttt{edges}$]
            \State $p_e \gets \Proc{ClosestPointOnEdge}(e,\ x)$
            \State $v \gets p_e - x$
            \State $d_e \gets |v|$
            \IF[$d_e < r$]
                \State $\texttt{hasTri}_0,\ n_0 \gets \Proc{GetAdjacentTriangleNormal}(e, 0)$
                \State $\texttt{hasTri}_1,\ n_1 \gets \Proc{GetAdjacentTriangleNormal}(e, 1)$
                \State $\texttt{isSilhouetteEdge} \gets \textbf{not}\ \texttt{hasTri}_0 \mid\mid \textbf{not}\ \texttt{hasTri}_1 \mid\mid$\\ $\hskip\algorithmicindent\hspace{39mm} (v \cdot n_0) \cdot (v \cdot n_1) \leq 0$
                \IFTHEN[$\texttt{isSilhouetteEdge}$]{$\textbf{return}\ d_e$}
                \ENDIFTHEN
            \ENDIF
        \ENDFORDO
    \ENDIF
    \ELSE
        \COMMENT[Intersect AABBs with sphere formed by $x$ and $r$]\ENDCOMMENT
        \State $L,R \gets T.\texttt{left}, T.\texttt{right}$
        \State $\texttt{visit}_L,\ d_L^{{\min}} \gets \Proc{IntersectAABBSphere}(L.\texttt{aabb},\ x,\ r)$
        \State $\texttt{visit}_R,\ d_R^{{\min}} \gets \Proc{IntersectAABBSphere}(R.\texttt{aabb},\ x,\ r)$

        \COMMENT[Cull nodes with only front- or back-facing triangles]\ENDCOMMENT
        \IF[$\texttt{visit}_L\ \textbf{and}\ d_L^{{\min}} > 0$]
            \State $\texttt{visit}_L \gets \Proc{HasSilhouette}(L.\texttt{aabb}, L.\texttt{cone}, x)$ \Comment{Alg. \ref{alg:HasSilhouette}}
        \ENDIF
        \IF[$\texttt{visit}_R\ \textbf{and}\ d_R^{{\min}} > 0$]
            \State $\texttt{visit}_R \gets \Proc{HasSilhouette}(R.\texttt{aabb}, R.\texttt{cone}, x)$ \Comment{Alg. \ref{alg:HasSilhouette}}
        \ENDIF
        \IF[$\texttt{visit}_L\ \textbf{and}\ \texttt{visit}_R$]            \IF[$d_L^{{\min}} < d_R^{{\min}}$] \Comment{Traverse closer subtree first}

                \State $r \gets \Proc{DistClosestSilhouette}(L,\ x,\ r,\ \smash{d_L^{{\min}}})$
                \State $\textbf{return}\ \Proc{DistClosestSilhouette}(R,\ x,\ r,\ \smash{d_R^{{\min}}})$
            \ENDIF
            \ELSE
                \State $r \gets \Proc{DistClosestSilhouette}(R,\ x,\ r,\ \smash{d_R^{{\min}}})$
                \State $\textbf{return}\ \Proc{DistClosestSilhouette}(L,\ x,\ r,\ \smash{d_L^{{\min}}})$
            \ENDELSE
        \ENDIF
        \ELSEIF[$\texttt{visit}_L$]
            \RETURN[$\Proc{DistClosestSilhouette}(L,\ x,\ r,\ \smash{d_L^{{\min}}})$]
            \ENDRETURN
        \ENDELSEIF
        \ELSEIF[$\texttt{visit}_R$]
            \RETURN[$\Proc{DistClosestSilhouette}(R,\ x,\ r,\ \smash{d_R^{{\min}}})$]
            \ENDRETURN
        \ENDELSEIF
    \ENDELSE
    \RETURN[$r$]
    \ENDRETURN
\end{algorithmic}
\end{algo}

\begin{algo}{\Proc{HasSilhouette}$(\texttt{aabb},\ \texttt{cone}_{\texttt{normal}},\ x)$}
\label{alg:HasSilhouette}
\begin{algorithmic}[1]
\algblockdefx[Name]{Class}{EndClass}
    [1][Unknown]{\textbf{class} #1}
    {}
\algtext*{EndClass}
\algblockdefx[Name]{FORDO}{ENDFORDO}
    [1][Unknown]{\textbf{for} #1 \textbf{do}}
    {}
\algtext*{ENDFORDO}
\algblockdefx[Name]{IF}{ENDIF}
    [1][Unknown]{\textbf{if} #1 \textbf{then}}
    {}
\algtext*{ENDIF}
\algblockdefx[Name]{IFTHEN}{ENDIFTHEN}
    [2][Unknown]{\textbf{if} #1 \textbf{then} #2}
    {}
\algtext*{ENDIFTHEN}
\algblockdefx[Name]{RETURN}{ENDRETURN}
    [1][Unknown]{\textbf{return} #1}
    {}
\algtext*{ENDRETURN}
\algblockdefx[Name]{COMMENT}{ENDCOMMENT}
    [1][Unknown]{\textcolor{commentpaleblue}{\(\triangleright\)#1}}
    {}
\algtext*{ENDCOMMENT}
\algblockdefx[Name]{ELSEIF}{ENDELSEIF}
    [1][Unknown]{\textbf{else if} #1 \textbf{then}}
    {}
\algtext*{ENDELSEIF}
\algblockdefx[Name]{ELSE}{ENDELSE}
    {\textbf{else}}
    {}
\algtext*{ENDELSE}

\Require An AABB $\texttt{aabb}$, a $\texttt{cone}_{\texttt{normal}}$ encoding normal information for the triangles in $\texttt{aabb}$, and a query point $x$.
\Ensure \emph{Conservative} guess of whether $\texttt{aabb}$ contains a silhouette with respect to $x$, achieved by checking if the normal and view cones associated with the AABB contain orthogonal directions.
    \COMMENT[$\texttt{aabb}$ may contain silhouette if $\texttt{cone}_{\texttt{normal}}$'s half angle is $\geq 90^{\circ}$]\ENDCOMMENT
    \State $a_{\texttt{nc}} \gets \texttt{cone}_{\texttt{normal}}.\texttt{axis}$
    \State $\theta_{\texttt{nc}} \gets \texttt{cone}_{\texttt{normal}}.\texttt{halfAngle}$
    \IFTHEN[$\theta_{\texttt{nc}} \geq \pi/2$]{$\textbf{return}\ \textsc{true}$}
    \ENDIFTHEN

    \COMMENT[Set view cone axis from $x$ to $\texttt{aabb}$'s center]\ENDCOMMENT
    \State $a_{\texttt{vc}} \gets \Proc{Centroid}($\texttt{aabb}$) - x$
    \State $l_{\texttt{vc}} \gets |a_{\texttt{vc}}|$
    \State $a_{\texttt{vc}} \gets a_{\texttt{vc}}\ /\ l_{\texttt{vc}}$\Comment{View cone axis}

    \COMMENT[$\texttt{aabb}$ may contain silhouette if $a_{\texttt{vc}}$ is $\perp$ to directions in $\texttt{cone}_{\texttt{normal}}$]\ENDCOMMENT
    \State $\phi \gets \Proc{ArcCos}(a_{\texttt{nc}} \cdot a_{\texttt{vc}})$\Comment{Angle between normal \&\ view axes}
    \IFTHEN[$\pi/2 \geq \phi\ -\ \theta_{\texttt{nc}}\ \textbf{and}\ \pi/2 \leq \phi\ +\ \theta_{\texttt{nc}}$]{$\textbf{return}\ \textsc{true}$}
    \ENDIFTHEN

    \COMMENT[Compute view cone half angle w.r.t. $\texttt{aabb}$'s bounding sphere]\ENDCOMMENT
    \State $r \gets \Proc{BoundingSphereRadius}(\texttt{aabb})$
    \IFTHEN[$l_{\texttt{vc}} \leq r$]{$\textbf{return}\ \textsc{true}$}\Comment{invalid cone, sphere contains $x$}
    \ENDIFTHEN
    \State $\theta_{\texttt{vc}} \gets \Proc{ArcSin}(r\ /\ l_{\texttt{vc}})$\Comment{View cone half angle}

    \COMMENT[$\texttt{aabb}$ may contain silhouette if cones contain $\perp$ directions]\ENDCOMMENT
    \State $\theta_{\texttt{sum}} \gets \theta_{\texttt{nc}}\ +\ \theta_{\texttt{vc}}$
    \IFTHEN[$\theta_{\texttt{sum}} \geq \pi/2$]{$\textbf{return}\ \textsc{true}$}
    \ENDIFTHEN
    \RETURN[$\pi/2 \geq \phi\ -\ \theta_{\texttt{sum}}\ \textbf{and}\ \pi/2 \leq \phi\ +\ \theta_{\texttt{sum}}$]
    \ENDRETURN
\end{algorithmic}
\end{algo}

\begin{algo}{\Proc{NeumannBoundarySample}$(x,\ r)$}
\label{alg:NeumannBoundarySample}
\begin{algorithmic}[1]
\algblockdefx[Name]{Class}{EndClass}
    [1][Unknown]{\textbf{class} #1}
    {}
\algtext*{EndClass}
\algblockdefx[Name]{FORDO}{ENDFORDO}
    [1][Unknown]{\textbf{for} #1 \textbf{do}}
    {}
\algtext*{ENDFORDO}
\algblockdefx[Name]{IF}{ENDIF}
    [1][Unknown]{\textbf{if} #1 \textbf{then}}
    {}
\algtext*{ENDIF}
\algblockdefx[Name]{IFTHEN}{ENDIFTHEN}
    [2][Unknown]{\textbf{if} #1 \textbf{then} #2}
    {}
\algtext*{ENDIFTHEN}
\algblockdefx[Name]{RETURN}{ENDRETURN}
    [1][Unknown]{\textbf{return} #1}
    {}
\algtext*{ENDRETURN}
\algblockdefx[Name]{COMMENT}{ENDCOMMENT}
    [1][Unknown]{\textcolor{commentpaleblue}{\(\triangleright\)#1}}
    {}
\algtext*{ENDCOMMENT}
\algblockdefx[Name]{ELSE}{ENDELSE}
    {\textbf{else}}
    {}
\algtext*{ENDELSE}

\Require A sphere with center $x \in \R^3$ and radius $r$.
   \Ensure A point $z \in \partial \domain_N$ if the sphere is nonempty, the unit outward normal \(n_z\) and the probability \(\texttt{pdf}_z\). The sample is not guaranteed to lie inside the sphere, but is likely to be near $x$.
    \COMMENT[Sample a random triangle inside or intersecting with the sphere]\ENDCOMMENT
    \State $t \gets \textsc{null}$
    \State $\texttt{pdf}_t \gets 0$
    \State $\Proc{SampleTriangleInSphere}(\texttt{snch}.\texttt{root},\ x,\ r,\ t,\ \texttt{pdf}_t)$
    \IF[$t\ \textbf{not}\ \textsc{null}$]
        \COMMENT[Sample a random point $z$ on the triangle $t$]\ENDCOMMENT
        \State $z,\ n_z,\ \texttt{pdf}_z \gets \Proc{SamplePointOnTriangle}(t)$
        \RETURN[$z,\ n_z,\ \texttt{pdf}_t \cdot \texttt{pdf}_z$]
        \ENDRETURN
    \ENDIF
    \RETURN[$\textsc{null},\ \textsc{null},\ 0$]
    \ENDRETURN
\end{algorithmic}
\end{algo}

\begin{algo}{\Proc{SampleTriangleInSphere}$(T,x,r,t,\texttt{pdf}_t,\texttt{pdf}_T=1)$}
\label{alg:SampleTriangleInSphere}
\begin{algorithmic}[1]
\algblockdefx[Name]{Class}{EndClass}
    [1][Unknown]{\textbf{class} #1}
    {}
\algtext*{EndClass}
\algblockdefx[Name]{FORDO}{ENDFORDO}
    [1][Unknown]{\textbf{for} #1 \textbf{do}}
    {}
\algtext*{ENDFORDO}
\algblockdefx[Name]{IF}{ENDIF}
    [1][Unknown]{\textbf{if} #1 \textbf{then}}
    {}
\algtext*{ENDIF}
\algblockdefx[Name]{IFTHEN}{ENDIFTHEN}
    [2][Unknown]{\textbf{if} #1 \textbf{then} #2}
    {}
\algtext*{ENDIFTHEN}
\algblockdefx[Name]{RETURN}{ENDRETURN}
    [1][Unknown]{\textbf{return} #1}
    {}
\algtext*{ENDRETURN}
\algblockdefx[Name]{COMMENT}{ENDCOMMENT}
    [1][Unknown]{\textcolor{commentpaleblue}{\(\triangleright\)#1}}
    {}
\algtext*{ENDCOMMENT}
\algblockdefx[Name]{ELSE}{ENDELSE}
    {\textbf{else}}
    {}
\algtext*{ENDELSE}

\Require A binary tree $T$, a sphere with center $x \in \R^3$ and radius $r$, a triangle $t$ yet to be selected and its sampling $\texttt{pdf}_t$. The optional argument computes $\texttt{pdf}_T$ of traversing a random branch in T.
\Ensure A randomly selected triangle $t$ in the sphere \& its sampling $\texttt{pdf}_t$; no triangle is selected if the sphere does not intersect $\partial \domain_N$.
    \IF[$T.\texttt{isLeaf}$]
        \COMMENT[Select a random triangle proportionally to its area]\ENDCOMMENT
        \State $\texttt{totalArea} \gets 0$
        \FORDO[$t_T\ \textbf{in}\ T.\texttt{triangles}$]
            \IF[$\Proc{IntersectTriangleSphere}(t_T,\ x,\ r)$]
                \State $\texttt{totalArea} \gets \texttt{totalArea}\ +\ \Proc{Area}(t_T)$
                \IF[$\Proc{Rand}() \cdot \texttt{totalArea} < \Proc{Area}(t_T)$]
                    \State $t \gets t_T$
                    \State $\texttt{pdf}_t \gets \texttt{pdf}_T \cdot \Proc{Area}(t)$
                \ENDIF
            \ENDIF
        \ENDFORDO
        \IFTHEN[$\texttt{totalArea} > 0$]{$\texttt{pdf}_t \gets \texttt{pdf}_t\ /\ \texttt{totalArea}$}
        \ENDIFTHEN
    \ENDIF
    \ELSE
        \COMMENT[Select subtree to traverse weighted by its proximity to $x$]\ENDCOMMENT
        \State $L \gets T.\texttt{left}$
        \State $R \gets T.\texttt{right}$
        \State $\texttt{weight}_L \gets \Proc{IntersectAABBSphere}(L.\text{aabb},\ x,\ r)\ ?$\\ $\hskip\algorithmicindent\hspace{19mm} \green^{\R^3}(x,\ \Proc{Centroid}(L.\text{aabb}))\ :\ 0$
        \State $\texttt{weight}_R \gets \Proc{IntersectAABBSphere}(R.\text{aabb},\ x,\ r)\ ?$\\ $\hskip\algorithmicindent\hspace{19mm} \green^{\R^3}(x,\ \Proc{Centroid}(R.\text{aabb}))\ :\ 0$
        \State $\texttt{totalWeight} \gets \texttt{weight}_L\ +\ \texttt{weight}_R$
        \IF[$\texttt{totalWeight} > 0$]
            \State $\texttt{prob}_L \gets \texttt{weight}_L\ /\ \texttt{totalWeight}$
            \IF[$\Proc{Rand}() < \texttt{prob}_L$]
                \State $\texttt{pdf}_L \gets \texttt{pdf}_T \cdot \texttt{prob}_L$
                \State $\Proc{SampleTriangleInSphere}(L,\ x,\ r,\ t,\ \texttt{pdf}_t,\ \texttt{pdf}_L)$
            \ENDIF
            \ELSE
                \State $\texttt{pdf}_R \gets \texttt{pdf}_T \cdot (1\ -\ \texttt{prob}_L)$
                \State $\Proc{SampleTriangleInSphere}(R,\ x,\ r,\ t,\ \texttt{pdf}_t,\ \texttt{pdf}_R)$
            \ENDELSE
        \ENDIF
    \ENDELSE
\end{algorithmic}
\end{algo}

\vfill

\newpage

\vfill

\end{document}